# Optimal design of auxetic hexachiral metamaterials with local resonators

Andrea Bacigalupo[1], Marco Lepidi[2], Giorgio Gnecco[1], Luigi Gambarotta[2]

[1]IMT Institute for Advanced Studies Lucca, Italy
[2]Department of Civil, Chemical and Environmental Engineering,
University of Genoa, Italy

**Abstract**

A parametric beam lattice model is formulated to analyse the propagation properties of elastic in-plane waves in an auxetic material based on a hexachiral topology of the periodic cell, equipped with inertial local resonators. The Floquet-Bloch boundary conditions are imposed on a reduced order linear model in the only dynamically active degrees-of-freedom. Since the resonators can be designed to open and shift band gaps, an optimal design, focused on the largest possible gap in the low-frequency range, is achieved by solving a maximization problem in the bounded space of the significant geometrical and mechanical parameters. A local optimized solution, for a the lowest pair of consecutive dispersion curves, is found by employing the globally convergent version of the Method of Moving asymptotes, combined with Monte Carlo and quasi-Monte Carlo multi-start techniques.

Corresponding author: Andrea Bacigalupo, mail: andrea.bacigalupo@imtlucca.it

# 1 Introduction

Auxetic materials possess the counterintuitive property of offering transversal expansions in response to longitudinal stretches. Such an unusual behaviour can be described, in the solid mechanics, by negative values of the Poisson ratios (Evans 1991, Lakes 1991, Evans and Alderson 2000, Alderson and Alderson 2007, Prawoto 2012). Although exceptionally experimented in natural materials, the artificial auxeticity has nowadays become a feasible manufactured achievement, by virtue of the recent extraordinary advances in the chemical engineering and structural micro-engineering fields. Indeed, negative Poisson ratios have been clearly documented in polymeric and metallic foams, mainly based on disordered open-cell bubble assemblies (Lakes 1987, Bianchi *et al.* 2008, Crichley *et al.* 2013), as well as in cellular periodic solids, primarily based on soft (possibly empty) matrices embedding ordered microstructures properly shaped as re-entrant honeycombs or rolling-up chiral patterns (Prall and Lakes 1997, Alderson *et al.* 2010, Spadoni and Ruzzene 2012, Dirrenberger *et al.* 2013).

The increasing research efforts towards the development of optimal design solutions, reliable mechanical models and efficient manufacture processes for auxetic materials are essentially motivated by a variety of challenging (and sometimes futuristic) applications in the chemical, marine, aerospace, nuclear, biomedical engineering fields, among the others. The reasons of such a growing success can be attributed to the high-performances of auxetic materials, not only in terms of non-conventional elastic properties, but also in respect to other functional qualities, including – for instance – augmented resistance to indentation and fracture, increased damping and effective passive filtering for vibrations.

With focus on micro-structured periodic materials, at least two major research issues can be recognized. On the one hand, a certain attention has been paid over the last years on defining homogeneous elastica (first and second order continua, Koiter and Cosserat continua), in which the overall elastic tensors are determined by means of standard or generalized macro-homogeneity conditions (Bazant and Christensen 1972, Kumar and McDowell 2004, Gonnella and Ruzzene 2008a, 2008b, Liu *et al.* 2012, Chen *et al.* 2013, Chen *et al.* 2014, Bacigalupo and Gambarotta 2014a, Bacigalupo and De Bellis 2015). On the other hand, an increasing interest has been recently attracted by the analysis of the transmission and dispersion properties of the elastic waves propagating across the material domain, either in its original periodic micro-structure (Phani et al. 2006, Tee et al. 2010, Spadoni *et al.* 2009, Bacigalupo and Lepidi 2015a) or in its equivalent homogenized form (Bacigalupo and Gambarotta 2014b, Bacigalupo and De Bellis 2015). Employing the Floquet-Bloch theory (Brillouin 1953), as it can be applied for instance to infinitely-periodic chains of elastically-coupled adjacent cells (Mead 1973, Romeo and Luongo 2002), several studies have been developed to parametrically assess the dispersion curves characterizing the wave frequency spectrum and, therefrom, the boundaries of frequency band-gaps lying between pairs of consecutive non-intersecting curves. In this respect, a promising improvement of the traditional chiral and antichiral systems, realized by a regular distribution of stiff disks/rings connected by flexible ligaments, consists in adding inter-ring massive inclusions, working as auxiliary oscillators elastically coupled to the microstructure (Liu *et al.* 2011, Tan *et al.* 2012, Bigoni *et al.* 2013, Bacigalupo and Gambarotta 2015). If properly tuned, these oscillators (*resonators*) may allow the adjustment and enhancement of the material performances, creating challenging perspectives in the optimal design of the frequency spectrum for specific purposes, such as opening, enlarging, closing or shifting band-gaps in target frequency ranges. Once completed, this achievement should potentially allow the realization of a novel class of fully customizable mechanical filters. In the practice, alternative feasible approaches consist in either seeking for an explicit, although approximate, parametric form of the dispersion curves (Craster *et. al.* 2010, Bacigalupo and Lepidi 2015b), suited to state an inverse eigenproblem, or formulating and solving an optimization problem, based on a suited objective function defined in the parameter space.

According to the latter solution, this paper employs a parametric, low-order dynamic model of the hexagonal unitary cell to state the linear eigenproblem governing the wave propagation in a hexachiral cellular material equipped with auxiliary resonators. Therefore, a set of design physical parameters is selected and their bounded range of technical values is employed as existence domain for seeking the solution of an optimization problem finalized to maximize the normalized angular frequency band-gap between two consecutive dispersion curves, particularly in the case of low frequencies. In the second part of the paper, the focus is on the formulation of such band gap optimization problem, the presentation of some approaches to solve it, and their numerical comparison.

The paper is organized as follows. Section 2 described the physical model. Section 3 formulates the band gap optimization problem and describes the main solution approach adopted. Section 4 reports and discusses the related numerical results, comparing the approach approach with another one. Extensions of the band gap optimization problem of Section 3 are presented in Section 4. Finally, Section 5 presents some conclusions. To make the paper self-consistent, some technical details are reported in Appendices A.1 and A.2.

## 2  Beam lattice model

A linear mechanical model is formulated to describe the free Hamiltonian dynamics of the micro-structured elementary cell tessellating a composite cellular material featured by a planar honeycomb geometry. The internal microstructure of each hexagonal cell, as well as the elastic coupling between adjacent cells, are determined by a periodic pattern of central rings connected to each other by six transcellular ligaments, spatially organised according to a hexachiral geometric topology (Figure 1a). From an intuitive perspective, the auxetic material behaviour can be physically justified by the particular ligament-ring arrangement, which tends to produce the same-sign iso-amplitude rotation of all the rings if the material is stretched along one of the ring alignments.

Focusing on the planar microstructure with unit thickness of the generic cell (Figure 1b), the central massive and highly-stiff ring is modelled as a rigid body, characterized by mean radius $R$ and width $w$. The light and highly-flexible ligaments are modelled as massless, linear, extensible, unshearable beams, characterized by natural length $L$ (between the ring-beam joints), transversal width $w$ and inclination $\beta$ (with respect to the line connecting the centres of adjacent rings of length $a$). By virtue of the periodic symmetry, the cell boundary crosses all the ligaments at midspan, halving their natural length. A heavy internal circular inclusion with external radius $r$, shown in Figure 1b (as a white circle), is located inside the ring through a soft elastic annulus (in grey). This inclusion, modelled as a rigid disk, plays the role of low-frequency (undamped) resonator.

The beam material is supposed linearly elastic, with Young's modulus $E_s$ and uniform mass density $\rho_s$ assigned to the ring material. The soft coating inside the resonator is considered as homogeneous, linearly elastic, isotropic material and has Young's modulus $E_r$ and Poisson's ratio $\nu_r$.

The major simplifying assumptions, introduced with the aim of reducing the dynamic model complexity without compromising its mechanical representativeness (according to Liu *et al.* 2011 and Bacigalupo and Gambarotta 2014a, 2014b, 2015), have a twofold justification. On the one hand, the rigidity of the rings and the disks is a technological requirement that can be easily accomplished. On the other hand, the neglection of the small inertia of the coating and ligaments can be considered acceptable in the present study, since their high-frequency natural vibrations are uninfluential on the analyses, finalized to the optimization of the low-frequency band gaps.

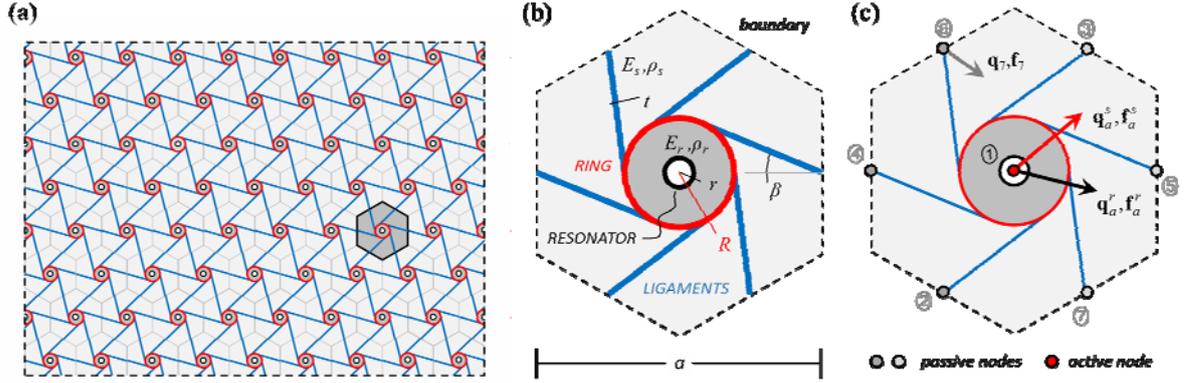

Figure 1: Hexachiral cellular material equipped with resonators: (a) pattern, (b) periodic hexagonal cell in the condition of maximal auxeticity (tangent ligaments, or $\beta = \arcsin(2R/a)$ ), (c) beam lattice model with nodal forces and displacements.

According to the mechanical assumptions and without further approximations related to the nodal discretization of the beam structure, a multi-degree-of-freedom lattice model, referred to a seven-node configuration for each cell, governs the undamped free dynamics of the periodic material. Focusing on the single cell model (Figure 1c), the generic $j$-th node (with $j = 1..7$) is employed as reference pole for three time-dependent generalized displacements of the structural elements, including the horizontal and vertical translations ($u_1^j, u_2^j$ for the beams/ring, $v_1, v_2$ for the resonator) and the in-plane rotation ($\varphi_j$ for the beams/ring, $\vartheta$ for the resonator). Depending on the specific mass distribution and with reference to Figure 1c, the full node set can be conveniently classified into

- a single-element subset composed of the only massive internal node (node 1), located at the coincident centroids of the ring and the resonator, which serves as common reference pole for all their *active* generalized displacements, collected in the six-by-one column-vector $\mathbf{q}_a = (\mathbf{q}_a^s, \mathbf{q}_a^r) = (u_1^1, u_2^1, \varphi^1, v_1, v_2, \vartheta)$, joining column-wise the ring subvector $\mathbf{q}_a^s = (u_1^1, u_2^1, \varphi^1)$ and the resonator subvector $\mathbf{q}_a^r = (v_1, v_2, \vartheta)$

- a six-element subset of massless external nodes (nodes 2-7), located at the midspan of the ligaments, whose *passive* displacements are collected in the 18-by-one column-vector $\mathbf{q}_p = (\mathbf{q}_2,\ldots,\mathbf{q}_7) = (u_1^2, u_2^2, \varphi^2, u_1^3, u_2^3, \varphi^3, \ldots, u_1^7, u_2^7, \varphi^7)$, casting column-wise the $j$-th node subvector $\mathbf{q}_j = (u_1^j, u_2^j, \varphi^j)$

The classification highlights that the local equilibrium of the active nodes is dynamically governed by the balance of elastic $\boldsymbol{\sigma}_a = (\boldsymbol{\sigma}_a^s, \boldsymbol{\sigma}_a^r)$ and inertial forces $\mathbf{f}_a = (\mathbf{f}_a^s, \mathbf{f}_a^r)$, whereas the local equilibrium of the passive nodes is quasi-statically established by the equality between the elastic forces $\boldsymbol{\sigma}_p$ and the reactive forces $\mathbf{f}_p$ transferred by the adjacent cells.

By virtue of the active/passive decomposition of the displacement and force vectors, the undamped free vibrations of the cell model are governed by the equilibrium equation

$$\begin{pmatrix} \mathbf{f}_a^s \\ \mathbf{f}_a^r \\ \mathbf{0} \end{pmatrix} + \begin{pmatrix} \boldsymbol{\sigma}_a^s \\ \boldsymbol{\sigma}_a^r \\ \boldsymbol{\sigma}_p \end{pmatrix} = \begin{pmatrix} \mathbf{0} \\ \mathbf{0} \\ \mathbf{f}_p \end{pmatrix}, \qquad (1)$$

where $\mathbf{f}_a^s$ and $\mathbf{f}_a^r$ are the inertial forces developed by the ring and the resonator, respectively. Making the inertial and elastic forces explicitly dependent on the respective accelerations and displacements, the equilibrium equation reads

$$\begin{bmatrix} \mathbf{M}^s & \mathbf{O} & \mathbf{O} \\ \mathbf{O} & \mathbf{M}^r & \mathbf{O} \\ \mathbf{O} & \mathbf{O} & \mathbf{O} \end{bmatrix} \begin{pmatrix} \ddot{\mathbf{q}}_a^s \\ \ddot{\mathbf{q}}_a^r \\ \ddot{\mathbf{q}}_p \end{pmatrix} + \begin{bmatrix} \mathbf{K}_{aa}^s & \mathbf{K}_{aa}^{sr} & \mathbf{K}_{ap}^s \\ \mathbf{K}_{aa}^{rs} & \mathbf{K}_{aa}^r & \mathbf{O} \\ \mathbf{K}_{pa}^s & \mathbf{O} & \mathbf{K}_{pp} \end{bmatrix} \begin{pmatrix} \mathbf{q}_a^s \\ \mathbf{q}_a^r \\ \mathbf{q}_p \end{pmatrix} = \begin{pmatrix} \mathbf{0} \\ \mathbf{0} \\ \mathbf{f}_p \end{pmatrix}, \qquad (2)$$

where dot indicates differentiation with respect to the time, while $\mathbf{O}$ stands for different-size empty matrices. The three-by-three positive definite diagonal submatrices $\mathbf{M}^s = \text{diag}(M_s, M_s, J_s)$ and $\mathbf{M}^r = \text{diag}(M_r, M_r, J_r)$ collect the translational and rotational masses of the ring ($M_s$ and $J_s$) and the resonator ($M_r$ and $J_r$), which depend on the structural geometry and the material properties as reported in Appendix A.1. The stiffness matrix in (2) is symmetric and positive definite. Its symmetric three-by-three submatrix $\mathbf{K}_{aa}^s$ (mainly) accounts for the stiffening effects of the six beams on the active displacements of the ring, whereas the 18-by-18 submatrix $\mathbf{K}_{pp}$ describes the stiffness of the passive external nodes. The coupling between the internal and external nodes (*global coupling*) is expressed by the submatrix $\mathbf{K}_{ap}^s = (\mathbf{K}_{pa}^s)^\mathrm{T}$. The resonator essentially behaves as a triad of independent linear (undamped) oscillators, each attached to the ring centroid by an elastic spring. The diagonal entries of the submatrix $\mathbf{K}_{aa}^r = \mathrm{diag}(k_d, k_d, k_\vartheta)$ define the equivalent spring Hookean constants, which depend on the structural geometry $R, r$ and the material properties $E_r, \nu_r$, namely $k_d = g_d(R, r, E_r, \nu_r)$ and $k_\vartheta = g_\vartheta(R, r, E_r, \nu_r)$ as reported in dimensionless form $k_d/E_s$ and $k_\vartheta/(a^2 E_s)$ in Appendix A.1. Consequently, the internal ring-resonator coupling (or *local coupling*) is expressed by the submatrix $\mathbf{K}_{aa}^{sr} = (\mathbf{K}_{aa}^{rs})^\mathrm{T} = -\mathbf{K}_{aa}^r$.

## 2.1 *Free wave propagation*

The free wave propagation along the cell domain spanned by $(x_1, x_2)$-coordinates of the central reference system can be studied according to the Floquet-Bloch theory (Brillouin 1953). Accordingly, the following representations of the *active* and *passive* generalized displacements and *passive* force vectors, in the $k$-transformed space, are introduced

$$\mathbf{q}_a = \mathbf{p}_a \exp(i\mathbf{k} \cdot \mathbf{x}_a), \qquad \mathbf{q}_p = \mathbf{F}_k \mathbf{p}_p, \qquad \mathbf{f}_p = \mathbf{F}_k \mathbf{b}_p, \qquad (3)$$

where $i^2 = -1$, $\mathbf{k} = (k_1, k_2)$ is the wave vector, $\mathbf{x}_a = (x_1^1, x_2^1)$ is the position vector of node 1, $\mathbf{p}_a = (\mathbf{p}_a^s, \mathbf{p}_a^r)$, $\mathbf{p}_p = (\mathbf{p}_2, \ldots, \mathbf{p}_7)$ and $\mathbf{b}_p = (\mathbf{b}_2, \ldots, \mathbf{b}_7)$ are *auxiliary* vectors and the 18-by-18 block diagonal matrix $\mathbf{F}_k$ reads

$$\mathbf{F}_k = \begin{bmatrix} \exp(i\mathbf{k} \cdot \mathbf{x}_2)\mathbf{I}_3 & 0 & 0 \\ 0 & \ddots & 0 \\ 0 & 0 & \exp(i\mathbf{k} \cdot \mathbf{x}_7)\mathbf{I}_3 \end{bmatrix}, \qquad (4)$$

being $\mathbf{x}_j = (x_1^j, x_2^j)$ the position vector of node $j$ (with $j = 1..7$) and $\mathbf{I}_3$ the three-by-three identity matrix.

First, the *passive* displacement/force vectors (and the *auxiliary passive* displacement/force vectors) can be conveniently decomposed as $\mathbf{q}_p = (\mathbf{q}_p^-, \mathbf{q}_p^+)$, $\mathbf{f}_p = (\mathbf{f}_p^-, \mathbf{f}_p^+)$ (and as $\mathbf{p}_p = (\mathbf{p}_p^-, \mathbf{p}_p^+)$, $\mathbf{b}_p = (\mathbf{b}_p^-, \mathbf{b}_p^+)$), in order to distinguish the nine nodal variables $\mathbf{q}_p^-, \mathbf{f}_p^-$ (and $\mathbf{p}_p^-, \mathbf{b}_p^-$) of the left boundary $\Gamma_-$ (composed by the external even nodes 2,4,6) from the nine nodal variables $\mathbf{q}_p^+, \mathbf{f}_p^+$ (and $\mathbf{p}_p^+, \mathbf{b}_p^+$) of the right boundary $\Gamma_+$ (composed by the external odd nodes 3,5,7). Therefore, the second and third equations of the (3) takes the following forms

$$\mathbf{q}_p^- = \mathbf{F}_k^- \mathbf{p}_p^-, \qquad \mathbf{q}_p^+ = \mathbf{F}_k^+ \mathbf{p}_p^+, \qquad \mathbf{f}_p^- = \mathbf{F}_k^- \mathbf{b}_p^-, \qquad \mathbf{f}_p^+ = \mathbf{F}_k^+ \mathbf{b}_p^+, \qquad (5)$$

where, assuming for the sake of convenience that the displacement and force vectors (and *auxiliary* displacement and force vectors) are sorted as $\mathbf{q}_p^+ = (\mathbf{q}_3, \mathbf{q}_5, \mathbf{q}_7)$, $\mathbf{q}_p^- = (\mathbf{q}_2, \mathbf{q}_4, \mathbf{q}_6)$ and $\mathbf{f}_p^+ = (\mathbf{f}_3, \mathbf{f}_5, \mathbf{f}_7)$, $\mathbf{f}_p^- = (\mathbf{f}_2, \mathbf{f}_4, \mathbf{f}_6)$ (and as $\mathbf{p}_p^+ = (\mathbf{p}_3, \mathbf{p}_5, \mathbf{p}_7)$, $\mathbf{p}_p^- = (\mathbf{p}_2, \mathbf{p}_4, \mathbf{p}_6)$ and $\mathbf{b}_p^+ = (\mathbf{b}_3, \mathbf{b}_5, \mathbf{b}_7)$, $\mathbf{b}_p^- = (\mathbf{b}_2, \mathbf{b}_4, \mathbf{b}_6)$), the 9-by-9 block diagonal matrices $\mathbf{F}_k^-$ and $\mathbf{F}_k^+$ reads

$$\mathbf{F}_k^- = \begin{bmatrix} \exp(i\mathbf{k}\cdot\mathbf{x}_2)\mathbf{I}_3 & 0 & 0 \\ 0 & \exp(i\mathbf{k}\cdot\mathbf{x}_4)\mathbf{I}_3 & 0 \\ 0 & 0 & \exp(i\mathbf{k}\cdot\mathbf{x}_6)\mathbf{I}_3 \end{bmatrix},$$

$$\mathbf{F}_k^+ = \begin{bmatrix} \exp(i\mathbf{k}\cdot\mathbf{x}_3)\mathbf{I}_3 & 0 & 0 \\ 0 & \exp(i\mathbf{k}\cdot\mathbf{x}_5)\mathbf{I}_3 & 0 \\ 0 & 0 & \exp(i\mathbf{k}\cdot\mathbf{x}_7)\mathbf{I}_3 \end{bmatrix}. \qquad (6)$$

Moreover, via the following periodic conditions on the *auxiliary* vectors

$$\mathbf{p}_p^+ = \mathbf{p}_p^-, \qquad \mathbf{b}_p^+ = -\mathbf{b}_p^-, \qquad (7)$$

the quasi-periodicity conditions enabling the free wave propagation throughout the cell domain between the two complementary boundaries, by recalling the equations (5) and (6), can be imposed by requiring

$$\mathbf{q}_p^+ = \mathbf{L}\mathbf{q}_p^-, \qquad \mathbf{f}_p^+ = -\mathbf{L}\mathbf{f}_p^-, \qquad (8)$$

where the nine-by-nine transfer matrix $\mathbf{L}$ reads

$$\mathbf{L} = \begin{bmatrix} \exp(i\mathbf{k}\cdot\mathbf{d}_{23})\mathbf{I}_3 & 0 & 0 \\ 0 & \exp(i\mathbf{k}\cdot\mathbf{d}_{45})\mathbf{I}_3 & 0 \\ 0 & 0 & \exp(i\mathbf{k}\cdot\mathbf{d}_{67})\mathbf{I}_3 \end{bmatrix}, \qquad (9)$$

where $\mathbf{k} = (k_1, k_2)$ is the wave vector and $\mathbf{d}_{ij} = \mathbf{x}_j - \mathbf{x}_i$ is the periodicity vector along the direction connecting the $\Gamma_-$-belonging *i*-th node to the $\Gamma_+$-belonging *j*-th node (Figure 2b in the dimensionless spaces).

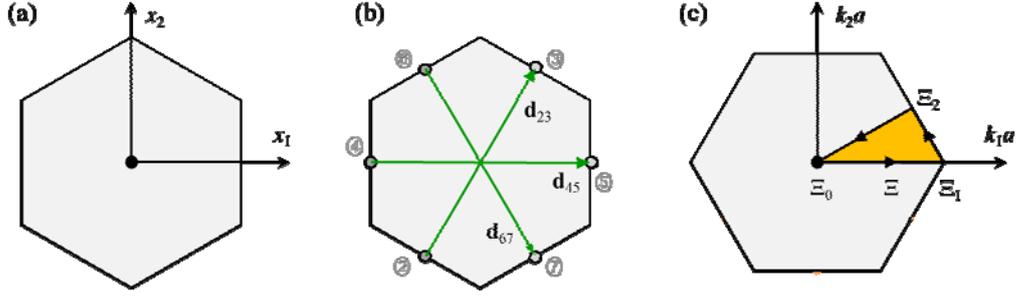

Figure 2: Periodic hexagonal cell: (a) central reference system, (b) periodicity vectors, (c) first irreducible Brillouin zone (with vertices $\bar{\mathbf{k}}_0 = (0,0)$, $\bar{\mathbf{k}}_1 = (0, 4/3\pi)$, $\bar{\mathbf{k}}_2 = (\pi, \sqrt{3}/3\,\pi)$, identified by the values of the curvilinear coordinate $\Xi$ equal to $\Xi_0, \Xi_1$, and $\Xi_2$, respectively).

Consistently with the passive displacement and force decomposition, and imposing the quasi-periodicity conditions (8), the lower (quasi-static) part of equation (2) reads

$$\begin{bmatrix} \mathbf{K}_{pa}^{s-} \\ \mathbf{K}_{pa}^{s+} \end{bmatrix} \mathbf{q}_a^s + \begin{bmatrix} \mathbf{K}_{pp}^{=} & \mathbf{K}_{pp}^{\mp} \\ \mathbf{K}_{pp}^{\pm} & \mathbf{K}_{pp}^{\#} \end{bmatrix} \begin{bmatrix} \mathbf{I}_9 \\ \mathbf{L} \end{bmatrix} \mathbf{q}_p^- = \begin{bmatrix} \mathbf{I}_9 \\ -\mathbf{L} \end{bmatrix} \mathbf{q}_p^-, \qquad (10)$$

where $\mathbf{I}_9$ stands for the nine-by-nine identity matrix. This equation can be solved to express the passive variables as *slave* functions of the *master* active displacements, yielding

$$\mathbf{q}_p^- = \mathbf{R}\left(\mathbf{K}_{pa}^{s+} + \mathbf{L}\mathbf{K}_{pa}^{s-}\right) \mathbf{q}_a^s, \qquad (11)$$

$$\mathbf{f}_p^- = \left(\mathbf{K}_{pa}^{s-} + \left(\mathbf{K}_{pp}^{\mp}\mathbf{L} + \mathbf{K}_{pp}^{=}\right)\mathbf{R}\left(\mathbf{K}_{pa}^{s+} + \mathbf{L}\mathbf{K}_{pa}^{s-}\right)\right) \mathbf{q}_a^s, \qquad (12)$$

where the auxiliary nine-by-nine matrix $\mathbf{R}$ is defined as $\mathbf{R} = -\left(\mathbf{L}\mathbf{K}_{pp}^{\mp}\mathbf{L} + \mathbf{L}\mathbf{K}_{pp}^{=} + \mathbf{K}_{pp}^{\pm} + \mathbf{K}_{pp}^{\#}\mathbf{L}\right)^{-1}$.

Similarly, the imposition of the quasi-periodicity conditions to the upper (dynamic) part of the equation (2), leads to the coupled equation

$$\begin{bmatrix} \mathbf{M}^r & \mathbf{O} \\ \mathbf{O} & \mathbf{M}^s \end{bmatrix} \begin{pmatrix} \ddot{\mathbf{q}}_a^s \\ \ddot{\mathbf{q}}_a^r \end{pmatrix} + \begin{bmatrix} \mathbf{K}_{aa}^s & \mathbf{K}_{aa}^{sr} \\ \mathbf{K}_{aa}^{rs} & \mathbf{K}_{aa}^r \end{bmatrix} \begin{pmatrix} \mathbf{q}_a^s \\ \mathbf{q}_a^r \end{pmatrix} + \begin{bmatrix} \mathbf{K}_{ap}^- & \mathbf{K}_{ap}^+ \\ \mathbf{O} & \mathbf{O} \end{bmatrix} \begin{bmatrix} \mathbf{I}_9 \\ \mathbf{L} \end{bmatrix} \mathbf{q}_p^- = \begin{pmatrix} \mathbf{0} \\ \mathbf{0} \end{pmatrix}, \qquad (13)$$

which can be decoupled by employing the master-slave relation (11) to quasi-statically condense the *auxiliary passive* displacements $\mathbf{q}_p^-$, yielding

$$\begin{bmatrix} \mathbf{M}^r & \mathbf{O} \\ \mathbf{O} & \mathbf{M}^s \end{bmatrix} \begin{pmatrix} \ddot{\mathbf{q}}_a^s \\ \ddot{\mathbf{q}}_a^r \end{pmatrix} + \begin{bmatrix} \hat{\mathbf{K}}_{aa}^s & \mathbf{K}_{aa}^{sr} \\ \mathbf{K}_{aa}^{rs} & \mathbf{K}_{aa}^r \end{bmatrix} \begin{pmatrix} \mathbf{q}_a^s \\ \mathbf{q}_a^r \end{pmatrix} = \begin{pmatrix} \mathbf{0} \\ \mathbf{0} \end{pmatrix}, \qquad (14)$$

being $\mathbf{K}_{aa}^{sr} = (\mathbf{K}_{aa}^{rs})^{\mathrm{T}} = -\mathbf{K}_{aa}^r$ and where the condensed submatrix $\hat{\mathbf{K}}_{aa}^s = \mathbf{K}_{aa}^s + \left(\mathbf{K}_{ap}^{s-} + \mathbf{K}_{ap}^{s+}\mathbf{L}\right)\mathbf{R}\left(\mathbf{K}_{pa}^{s+} + \mathbf{L}\mathbf{K}_{pa}^{s-}\right)$ and the full stiffness matrix can be proved to be Hermitian.

Finally, imposing the $\omega$-angular frequency harmonically oscillating solution for all the active variables $\mathbf{q}_a^s$ and $\mathbf{q}_a^r$, i.e.

$$\begin{aligned} \mathbf{q}_a^s &= \mathbf{p}_a^s \exp(i\mathbf{k}\cdot\mathbf{x}_a) = \boldsymbol{\psi}_a^s \exp(i\omega t)\exp(i\mathbf{k}\cdot\mathbf{x}_a), \\ \mathbf{q}_a^r &= \mathbf{p}_a^r \exp(i\mathbf{k}\cdot\mathbf{x}_a) = \boldsymbol{\psi}_a^r \exp(i\omega t)\exp(i\mathbf{k}\cdot\mathbf{x}_a), \end{aligned} \qquad (15)$$

with $i^2 = -1$, eliminating the $t$ time-dependence and posing $\lambda = \omega^2$, a linear eigenproblem is stated in the non-standard form

$$\left( \begin{bmatrix} \hat{\mathbf{K}}_{aa}^s & \mathbf{K}_{aa}^{sr} \\ \mathbf{K}_{aa}^{rs} & \mathbf{K}_{aa}^r \end{bmatrix} - \lambda \begin{bmatrix} \mathbf{M}^s & \mathbf{O} \\ \mathbf{O} & \mathbf{M}^r \end{bmatrix} \right) \begin{pmatrix} \boldsymbol{\psi}_a^s \\ \boldsymbol{\psi}_a^r \end{pmatrix} = \begin{pmatrix} \mathbf{0} \\ \mathbf{0} \end{pmatrix}, \qquad (16)$$

whose *eigensolutions* are composed by six eigenvalues $\lambda \in \mathbb{R}$ and the corresponding six complex-valued eigenvectors $\boldsymbol{\psi}_a = \left(\boldsymbol{\psi}_a^s, \boldsymbol{\psi}_a^r\right) \in \mathbb{C}^6$ in the $\omega, k$-transformed space, including six active eigencomponents each. The passive eigencomponents can be determined by applying the quasi-static relation $\boldsymbol{\psi}_p^- = \mathbf{R}\left(\mathbf{K}_{pa}^{s+} + \mathbf{L}\mathbf{K}_{pa}^{s-}\right)\boldsymbol{\psi}_a^s$ and the quasi periodicity condition $\boldsymbol{\psi}_p^+ = \mathbf{L}\boldsymbol{\psi}_p^-$. In particular, each eigenvectors $\boldsymbol{\psi}_a$ is the polarization vector of the plane harmonic wave travelling along $\mathbf{k}$ with angular frequency $\omega$.

Therefore, the wave propagation in the cellular material can be fully characterized by following the eigenvalues $\lambda$ under variation of the wave vector $\mathbf{k}$ in the irreducible Brillouin *B*-range analogous (Brillouin 1953) to the one illustrated in Figure 2c. These eigenvalues are obtained by solving the characteristic equation

$$\det\left(\tilde{\mathbf{K}} - \lambda\tilde{\mathbf{M}}\right) = 0, \qquad (17)$$

associated to equation (16) where the two matrices are

$$\tilde{\mathbf{K}} = \begin{bmatrix} \hat{\mathbf{K}}_{aa}^s & \mathbf{K}_{aa}^{sr} \\ \mathbf{K}_{aa}^{rs} & \mathbf{K}_{aa}^r \end{bmatrix}, \qquad \tilde{\mathbf{M}} = \begin{bmatrix} \mathbf{M}^s & \mathbf{O} \\ \mathbf{O} & \mathbf{M}^r \end{bmatrix}, \qquad (18)$$

where it is worth recalling that $\mathbf{K}_{aa}^{sr} = (\mathbf{K}_{aa}^{rs})^{\mathrm{T}} = -\mathbf{K}_{aa}^{r}$.

Since the ligament natural length $L$ is a ($a, \beta, R$)-dependent parameter obeying to the relation $L = a\left(\cos\beta - \sqrt{(2R/a)^2 - \sin^2\beta}\right)$, the physical properties of the cell microstructure are completely defined by assessing the following set of parameters

$$\boldsymbol{\mu} = \left(a, w, R, \beta, r, E_s, \rho_s, \nu_r, E_r, \rho_r\right). \tag{19}$$

In particular, a triangular beam lattice is obtained for the limit case $\beta \to 0$ and for this geometry the microstructure is no longer chiral.

In terms of the dimensionless wave vector $\bar{\mathbf{k}} = a\mathbf{k}$, the dimensionless eigenvalues $\bar{\lambda} = \dfrac{\lambda a^2 \rho_s}{E_s}$, the dimensionless angular frequency $\bar{\omega} = \dfrac{\omega a}{\sqrt{E_s/\rho_s}}$ and the minimal set of seven independent dimensionless parameters (see Appendix A.1)

$$\bar{\boldsymbol{\mu}} = \left(\frac{w}{a}, \frac{R}{a}, \beta, \frac{r}{a}, \nu_r, \frac{E_r}{E_s}, \frac{\rho_r}{\rho_s}\right) \tag{20}$$

the linear eigenproblem (16) takes the following equivalent dimensionless form

$$\left(\bar{\mathbf{K}}(\bar{\boldsymbol{\mu}}, \bar{\mathbf{k}}) - \bar{\lambda}\bar{\mathbf{M}}(\bar{\boldsymbol{\mu}})\right)\bar{\boldsymbol{\psi}}_a = \mathbf{0}, \tag{21}$$

being $\bar{\boldsymbol{\psi}}_a$ the dimensionless eigenvector. Its characteristic equation is

$$\det\left(\bar{\mathbf{K}}(\bar{\boldsymbol{\mu}}, \bar{\mathbf{k}}) - \bar{\lambda}\bar{\mathbf{M}}(\bar{\boldsymbol{\mu}})\right) = 0, \tag{22}$$

where the $(\bar{\boldsymbol{\mu}}, \bar{\mathbf{k}})$- dimensionless dependent matrices are

$$\bar{\mathbf{K}}(\bar{\boldsymbol{\mu}}, \bar{\mathbf{k}}) = \begin{bmatrix} \hat{\bar{\mathbf{K}}}_{aa}^{s} & \bar{\mathbf{K}}_{aa}^{sr} \\ \bar{\mathbf{K}}_{aa}^{rs} & \bar{\mathbf{K}}_{aa}^{r} \end{bmatrix}, \quad \bar{\mathbf{M}}(\bar{\boldsymbol{\mu}}) = \begin{bmatrix} \bar{\mathbf{M}}^{s} & \mathbf{O} \\ \mathbf{O} & \bar{\mathbf{M}}^{r} \end{bmatrix}, \tag{23}$$

with $\bar{\mathbf{K}}_{aa}^{sr} = (\bar{\mathbf{K}}_{aa}^{rs})^{\mathrm{T}} = -\bar{\mathbf{K}}_{aa}^{r}$ and their components are extensively reported in Appendix A.2. For fixed $\bar{\boldsymbol{\mu}}$, the $i$-th dimensionless angular frequency locus $\bar{\omega}_i\left(\bar{\boldsymbol{\mu}}, \bar{\mathbf{k}}(\Xi)\right)$ along the closed boundary $\partial B$ of the Brillouin $B$-range, spanned anticlockwise by the dimensionless curvilinear coordinate $\Xi$ (shown in Figure 2c), is the $i$-th dispersion curve of the Floquet-Bloch spectrum. In particular, the $\partial B$-vertices (of the first irreducible Brillouin zone) are scanned in the order $\bar{\mathbf{k}}_0 = (0,0)$, $\bar{\mathbf{k}}_1 = (0, 4/3\pi)$, $\bar{\mathbf{k}}_2 = (\pi, \sqrt{3}/3\ \pi)$, $\bar{\mathbf{k}}_0 = (0,0)$ and are identify of the increasing $\Xi$-values $\Xi_0 = 0$, $\Xi_1 = \dfrac{4}{3}\pi$, $\Xi_2 = 2\pi$ and $\Xi_0 = 2\left(1+\sqrt{3}/3\right)\pi$, respectively.

## 3 Band gap optimization problem

A research issue of major theoretical and applied interest in the rapidly-evolving field of engineered composite materials consists in the detection, quantification and – as a final target – design of the normalized angular frequency band-gap between two consecutive dispersion curves (see, e.g., Ruzzene and Scarpa 2005, in which a periodic auxetic lattice was optimized with respect to one design parameter). Mostly oriented to strongly promising applications of these materials as fully mechanical filters, the strongest research efforts are currently devoted to the fine adjustment of the design parameters in order to maximize the amplitude of the low-frequency band-gaps.

### 3.1 *Problem formulation*

The maximization of the normalized angular frequency band gap between a pair of consecutive dispersion curves can be based on the definition of a suited $\bar{\boldsymbol{\mu}}$-dependent objective function $\Delta\bar{\omega}_{hk}(\bar{\boldsymbol{\mu}})$, and thus formulated as a constrained optimization problem:

$$\begin{aligned}
&\underset{\bar{\boldsymbol{\mu}}}{\text{maximize}}\, \Delta\bar{\omega}_{hk}(\bar{\boldsymbol{\mu}}) \\
&\text{subject to } \bar{\mu}_i^{\min} \leq \bar{\mu}_i \leq \bar{\mu}_i^{\max}, \ i=1,\ldots,d,\ i\neq 3,4 \\
&\qquad\qquad \bar{\mu}_i^{\min} \leq \bar{\mu}_i \leq \bar{\mu}_i^{\max}(\bar{\mu}_2),\ i=3 \\
&\qquad\qquad \bar{\mu}_i^{\min}(\bar{\mu}_2) \leq \bar{\mu}_i \leq \bar{\mu}_i^{\max}(\bar{\mu}_2),\ i=4
\end{aligned} \qquad (24)$$

where $d=7$, $\bar{\boldsymbol{\mu}} \in \mathbb{R}^d$ is the vector of dimensionless design parameters $\bar{\mu}_i$ (with $i=1..d$) to be optimally assessed, $\bar{\mu}_i^{\min}$ (which is $\bar{\mu}_i^{\min}(\bar{\mu}_2)$ for $i=4$) is a lower bound on the design variable $\bar{\mu}_i$, and $\bar{\mu}_i^{\max}$ (which is $\bar{\mu}_i^{\max}(\bar{\mu}_2)$ for $i=3,4$) is an upper bound on the design variable $\bar{\mu}_i$. The lower/upper bounds on each parameter $\bar{\mu}_i$ are established in Table 1, according to geometrical, physical or technological requirements.

Recalling that $\bar{\mathbf{k}} \in \mathbb{R}^2$ is the dimensionless wave vector, sorting the dimensionless angular eigenfrequencies $\bar{\omega}_i$ in ascending order and denoting $\partial B_H$ a uniform discretization (with $H$ equispaced points) of the boundary of the first irreducible Brillouin zone (see Figure 2c), the objective function is defined as

$$\Delta \bar{\omega}_{hk}(\bar{\mu}) = \min_{\bar{\mathbf{k}} \in \partial B_H} \bar{\omega}_h(\bar{\mu}, \bar{\mathbf{k}}) - \max_{\bar{\mathbf{k}} \in \partial B_H} \bar{\omega}_k(\bar{\mu}, \bar{\mathbf{k}}), \qquad (25)$$

measuring the positive (even if possibly null, or even negative) band-gap amplitude $\Delta \bar{\omega}_{hk}(\bar{\mu}, \bar{\mathbf{k}})$ between the $h$-th and $k$-th consecutive dispersion curves (where $h = k+1$). Finally, the bound on the design parameters to be optimized are reported in Table 1.

Table 1: lower and upper bounds on the optimizable parameters

| $\bar{\mu}_i$ | $w/a$ | $R/a$ | $\beta$ | $r/a$ | $v_a$ | $E_r/E_s$ | $\rho_r/\rho_s$ |
|---|---|---|---|---|---|---|---|
| $\bar{\mu}_i^{\min}$ | $\dfrac{3}{50}$ | $\dfrac{1}{10}$ | $0$ | $\dfrac{1}{2}\dfrac{R}{a}$ | $\dfrac{2}{10}$ | $\dfrac{1}{10}$ | $\dfrac{1}{10}$ |
| $\bar{\mu}_i^{\max}$ | $\dfrac{1}{10}$ | $\dfrac{1}{5}$ | $\arcsin\left(2\dfrac{R}{a}\right)$ | $\dfrac{9}{10}\dfrac{R}{a}$ | $\dfrac{4}{10}$ | $10$ | $10$ |

Although the problem can be specified for any pair of consecutive dispersion curves, it is desirable to seek maxima of the objective function in the lowest possible spectrum band, for which a positive band gap exist between the corresponding dispersion curves, for suitable choices of the parameters. As far as the *bounded* variation of the design parameters does not alter the centro-symmetric geometry, all the eigenvalues are verified to attain positive values ($\lambda \in \mathbb{R}^+$), corresponding to real-valued positive frequencies. Furthermore, the lowest frequency pair is systematically found to coincide and vanish ($\omega_1 = \omega_2 = 0$) for $\bar{\mathbf{k}} = \mathbf{0}$, independently of the particular $\bar{\mu}$-assignment within the established bounds (Bacigalupo and Gambarotta 2015). Consequently, stating the optimization problem for $h = 2, k = 1$ (that is, adopting the function $\Delta \bar{\omega}_{21}(\bar{\mu})$) turns out useless, because the first and second dispersion curves are expected to intersect to each other at least in the origin $\bar{\mathbf{k}} = \mathbf{0}$. Therefore, the amplitude of the lowest frequency band-gap $\Delta \bar{\omega}_{32}(\bar{\mu})$ (setting $h = 3$, $k = 2$) is investigated in the following, where $H = 30$ to have a sufficiently dense investigation range.

It is clear from formulas (21)-(23) and the expressions of the matrices $\bar{\mathbf{K}}(\bar{\mu}, \bar{\mathbf{k}})$ and $\overline{\mathbf{M}}(\overline{\boldsymbol{\mu}})$, reported in Appendix A.2, that the band gap optimization problem (24) is a nonlinear programming problem. Moreover, its objective function $\Delta\overline{\omega}_{hk}(\overline{\boldsymbol{\mu}})$ is not a concave function. Hence, the optimization problem (24) cannot be treated as a concave maximization problem, and this makes challenging finding its globally optimal solution. Therefore, a locally optimal solution is sought in the following.

### 3.2 *Solution with GCMMA optimization method*

To solve locally the band gap optimization problem for the auxetic hexachiral structure, the Globally Convergent version of the Method of Moving Asymptotes (GCMMA) (Svanberg 2002) is exploited in the following. Such a method has been often used in the structural optimization literature to solve similar band gap optimization problems, for both phononic and photonic structures (Sigmund 2003 and Diaz *et al.* 2004), since such problems have a similar mathematical formulation. Then, to improve the quality of the obtained solution, in the paper the method is also combined with suitable multi-start techniques (which allow to apply the method multiple times, starting from different suitably selected initializations). In the following, a short description of both is provided.

GCMMA is an extension of the Method of Moving Asymptotes (MMA) (Svanberg 1987) which searches for a locally optimal solution of a nonlinear programming problem by solving a sequence of simpler maximization sub-problems, at each iteration $m$. These are obtained by approximating the objective and constraint functions of the original optimization problem around the current vector $\overline{\boldsymbol{\mu}}^{(m)}$ of design variables, and updating such variables after solving each sub-problem. In particular, the focus here is on its application to the optimization problem (24), a case for which each sub-problem approximates the functions $\Delta\overline{\omega}_{hk}(\overline{\boldsymbol{\mu}})$, $\overline{\mu}_i^{\min}(\overline{\mu}_2)$ for $i=4$, and $\overline{\mu}_i^{\max}(\overline{\mu}_2)$ for $i=3,4$. From an optimization perspective, each sub-problem has the following nice properties (see, e.g., Christensen and Klarbring 2004):

1) the approximations $\Delta\widetilde{\widetilde{\omega}}_{hk}^{(m)}(\overline{\boldsymbol{\mu}})$, $\widetilde{\widetilde{\mu}}_i^{\min,(m)}(\overline{\mu}_2)$, and $\widetilde{\widetilde{\mu}}_i^{\max,(m)}(\overline{\mu}_2)$ of $\Delta\overline{\omega}_{hk}(\overline{\boldsymbol{\mu}})$, $\overline{\mu}_i^{\min}(\overline{\mu}_2)$, and $\overline{\mu}_i^{\max}(\overline{\mu}_2)$, respectively, are first-order approximations, in the sense that, for all these functions, when the function to be approximated is locally differentiable, there is no error in the approximation of the function value and of its gradient when evaluated at the current design variables (for the objective function, local non-differentiability may occur in case the band gap at the current design variables is 0, if this is due to the second and third dispersion curves being tangent at one point of the domain);

2) such approximations are concave functions;

3) the approximation $\Delta \widetilde{\overline{\omega}}_{hk}^{(m)}(\overline{\boldsymbol{\mu}})$ is separable, in the sense they it is the sum of functions of one variable (one function for each design variable), which makes each optimization sub-problem quite easy to solve through standard Lagrange multiplier techniques (of course, the approximations $\widetilde{\overline{\mu}}_i^{\min,(m)}(\overline{\mu}_2)$ and $\widetilde{\overline{\mu}}_i^{\max,(m)}(\overline{\mu}_2)$ are separable by definition, since $\overline{\mu}_2$ is a single design variable).

Besides MMA, the properties 1), 2), and 3) above are satisfied also by other optimization methods that are often used in structural optimization, such as sequential linear programming and convex linearization (CONLIN) (Christensen and Klarbring 2004). The difference is that MMA is based on a more flexible approximation, which is generated using a technique named of "moving asymptotes". This means, e.g., that each approximation $\Delta \widetilde{\overline{\omega}}_{hk}^{(m)}(\overline{\boldsymbol{\mu}})$ has the form

$$\Delta \widetilde{\overline{\omega}}_{hk}^{(m)}(\overline{\boldsymbol{\mu}}) = \Delta \overline{\omega}_{hk}(\overline{\boldsymbol{\mu}}^{(m)}) + \sum_{i=1}^{d} \frac{\Delta \widetilde{\overline{\omega}}_{hk}^{U_i,(m)}}{U_i^{(m)} - \overline{\mu}_i} + \sum_{i=1}^{d} \frac{\Delta \widetilde{\overline{\omega}}_{hk}^{L_i,(m)}}{\overline{\mu}_i - L_i^{(m)}}, \tag{26}$$

where, for each iteration, $\Delta \widetilde{\overline{\omega}}_{hk}^{U_i,(m)}$, $\Delta \widetilde{\overline{\omega}}_{hk}^{L_i,(m)}$, $U_i^{(m)}$ and $L_i^{(m)}$ are suitable constants (see Svanberg 1987), and, to get a bounded approximation, one also adds the constraints

$$L_i^{(m)} < a_i^{(m)} \leq \overline{\mu}_i^{(m)} \leq b_i^{(m)} < U_i^{(m)}, \tag{27}$$

for other suitable constants $a_i^{(m)}$ and $b_i^{(m)}$ (see Svanberg 1987). The name of the method derives from the fact that the vertical lines $\overline{\mu}_i = L_i^{(m)}$ and $\overline{\mu}_i = U_i^{(m)}$ are asymptotes for the approximation (26), which *move* (i.e., they are updated in a suitable way) from each iteration to the successive one. It is worth mentioning that, among other reasons, the larger flexibility of MMA with respect to sequential linear programming and CONLIN is also motivated by the fact that both are obtained as limit cases of MMA, for limit choices of its asymptotes (Christensen and Klarbring 2004). It is also worth remarking that MMA may not always converge to a stationary point of the original optimization problem. For this reason, its variation GCMMA was presented in Svanberg 2002 as a globally convergent version of MMA, in which the convergence of the modified method to a stationary point of the original problem is guaranteed. However, due to the high nonlinearity of that problem, such a point is not guaranteed to be its global minimizer.

In order to improve the quality of the solution obtained by GCMMA, in the following such a method is combined with a multi-start technique. This means that the method is applied repeatedly a number $S$ of times, with different initializations, and the best design vector found in all the repetitions is taken as a surrogate of a globally optimal design vector. As for the specific choice of the multi-start technique, the following two approaches are considered in the following:

1) a Monte Carlo initialization of the design variables, with $\bar{\mu}_i$, for $i \neq 3,4$, taken as realizations of independent uniformly distributed random variables with supports $\left[\bar{\mu}_i^{\min}, \bar{\mu}_i^{\max}\right]$, respectively, and $\bar{\mu}_3, \bar{\mu}_4$ sampled subsequently and independently according to two uniform probability distributions with supports $\left[\bar{\mu}_3^{\min}, \bar{\mu}_3^{\max}(\bar{\mu}_2)\right]$ and $\left[\bar{\mu}_4^{\min}(\bar{\mu}_2), \bar{\mu}_4^{\max}(\bar{\mu}_2)\right]$, respectively;

2) a quasi-Monte Carlo initialization, obtained at first generating a quasi-random Sobol' sequence (see, e.g., Niederreiter 1992 for its definition) on the 7-dimensional unit cube $[0,1]^7$, then, applying to every vector belonging to such a sequence the mapping $\mathbf{h}:[0,1]^7 \to \mathbb{R}^7$, with $\mathbf{h}(\mathbf{y})$ being defined as

$$\mathbf{h}(\mathbf{y}) = \begin{pmatrix} \bar{\mu}_1(\mathbf{y}) \\ \bar{\mu}_2(\mathbf{y}) \\ \bar{\mu}_3(\mathbf{y}) \\ \bar{\mu}_4(\mathbf{y}) \\ \bar{\mu}_5(\mathbf{y}) \\ \bar{\mu}_6(\mathbf{y}) \\ \bar{\mu}_7(\mathbf{y}) \end{pmatrix} = \begin{pmatrix} \bar{\mu}_1^{\min} + (\bar{\mu}_1^{\max} - \bar{\mu}_1^{\min})y_1 \\ \bar{\mu}_2^{\min} + (\bar{\mu}_2^{\max} - \bar{\mu}_2^{\min})y_2 \\ \bar{\mu}_3^{\min} + (\bar{\mu}_3^{\max}(\bar{\mu}_2(y_2)) - \bar{\mu}_3^{\min})y_3 \\ \bar{\mu}_4^{\min} + (\bar{\mu}_4^{\max}(\bar{\mu}_2(y_2)) - \bar{\mu}_4^{\min}(\bar{\mu}_2(y_2)))y_4 \\ \bar{\mu}_5^{\min} + (\bar{\mu}_5^{\max} - \bar{\mu}_5^{\min})y_5 \\ \bar{\mu}_6^{\min} + (\bar{\mu}_6^{\max} - \bar{\mu}_6^{\min})y_6 \\ \bar{\mu}_7^{\min} + (\bar{\mu}_7^{\max} - \bar{\mu}_7^{\min})y_7 \end{pmatrix}, \quad (28)$$

Compared with the Monte Carlo initialization method, the quasi-Monte Carlo approach has the advantages of being exactly replicable, and of generating more *uniform* sequences of initial points, whereas with the Monte Carlo approach there is in principle the possibility of generating the same initial point (or very similar initial points) more than once in the sequence.

In the following, the combination of the GCMMA algorithm with the Monte Carlo initialization 1) and the quasi-Monte Carlo initialization 2) are referred, respectively, as GCMMA/MC and GCMMA/QMC.

## 4 Results and discussion

The following Subsection 4.1 reports the results of the comparison between the GCMMA algorithm combined with the two initialization methods, and a less-sophisticated (and less efficient) optimization approach, known in the literature as brute-force approach. It is shown therein that all the methods are able to find similar solutions, but with GCMMA/MC and GCMMA/QMC requiring a much smaller number of objective function evaluations, and a larger best value of the objective function than the brute-force approach. Then, Subsection 4.2 investigates in more detail the best solution obtained by all the methods (which is the one found by GCMMA/QMC), analysing the coupling between the ring and the resonator for the associated values of the design parameters.

4.1 *Comparison of GCMMA/MC and GCMMA/QMC with the brute-force approach*

In order to assess the quality of the solution produced by the combination of the GMMA algorithm with the multi-start technique, in the following a comparison with the brute-force approach is reported. Applied to the present context, this approach is characterized by the fact that the design variables $\bar{\mu}_i$, for $i \neq 3,4$, are uniformly discretized on $\left[\bar{\mu}_i^{\min}, \bar{\mu}_i^{\max}\right]$ using a number $N_i$ of discretization levels each, then $\bar{\mu}_3$ and $\bar{\mu}_4$ are uniformly discretized on $\left[\bar{\mu}_3^{\min}, \bar{\mu}_3^{\max}(\bar{\mu}_2)\right]$ and $\left[\bar{\mu}_4^{\min}(\bar{\mu}_2), \bar{\mu}_4^{\max}(\bar{\mu}_2)\right]$, respectively, using $N_3$ and $N_4$ discretization levels (here, the discretization step, i.e., the distance between two consecutive levels, depends on the choice of $\bar{\mu}_2$). Finally, the band gap is evaluated and maximized (by explicit enumeration) on the resulting grid of vectors of design parameters. Of course, such an approach cannot be practically extended to problems with a sufficiently larger number of design variables, and has to be limited, in any case, to coarse grids, as being subject to the *curse of dimensionality* (see Bellman 1957). In other words, in the general case, being $d$ the number of design variables and $N_i$ the number of discretization levels for each of them, the brute-force approach requires $\prod_{i=1}^{d} N_i$ evaluations of the band gap (one for each choice of the set of discretized design variables, forming the optimization grid), which is exponential in the number of design variables when $N_i = N$ for all $i$. Nevertheless, for an optimization problem of small size (e.g., with a small number of optimization variables), simple constraints, and a sufficiently smooth objective function (like problem (24)), the brute-force approach can provide a good approximation of the optimal value of the objective function (even though, likely, with a computational effort larger than other methods), hence providing a benchmark for the evaluation of other more efficient algorithms.

First of all, the brute-force approach provides a motivation for the introduction of the resonator to obtain a positive band gap. Without the resonator, indeed, the brute-force approach makes it possible to obtain a very good approximation of the optimal value of the objective, since the number of design variables is reduced to three, i.e., $\frac{w}{a}$, $\frac{R}{a}$ and $\beta$ (the other four design variables are set to zero, and all the constraints for $i = 4,5,6,7$ are removed from the problem formulation (24)), and a fine discretization is possible with a reasonably small computational burden. Then, by selecting $N_i = 10$ discretization steps for $i = 1,2,3$, the objective function has been evaluated for $10^3 = 1000$ different choices of this *reduced* design vector, obtaining zero as its largest value. This, combined with the smoothness of the objective function (illustrated in Figure 3, which plots the objective as a function of $\frac{w}{a}$ and $\beta$, for several values of $\frac{R}{a}$) allows to conclude that without the resonator there is no positive band gap.

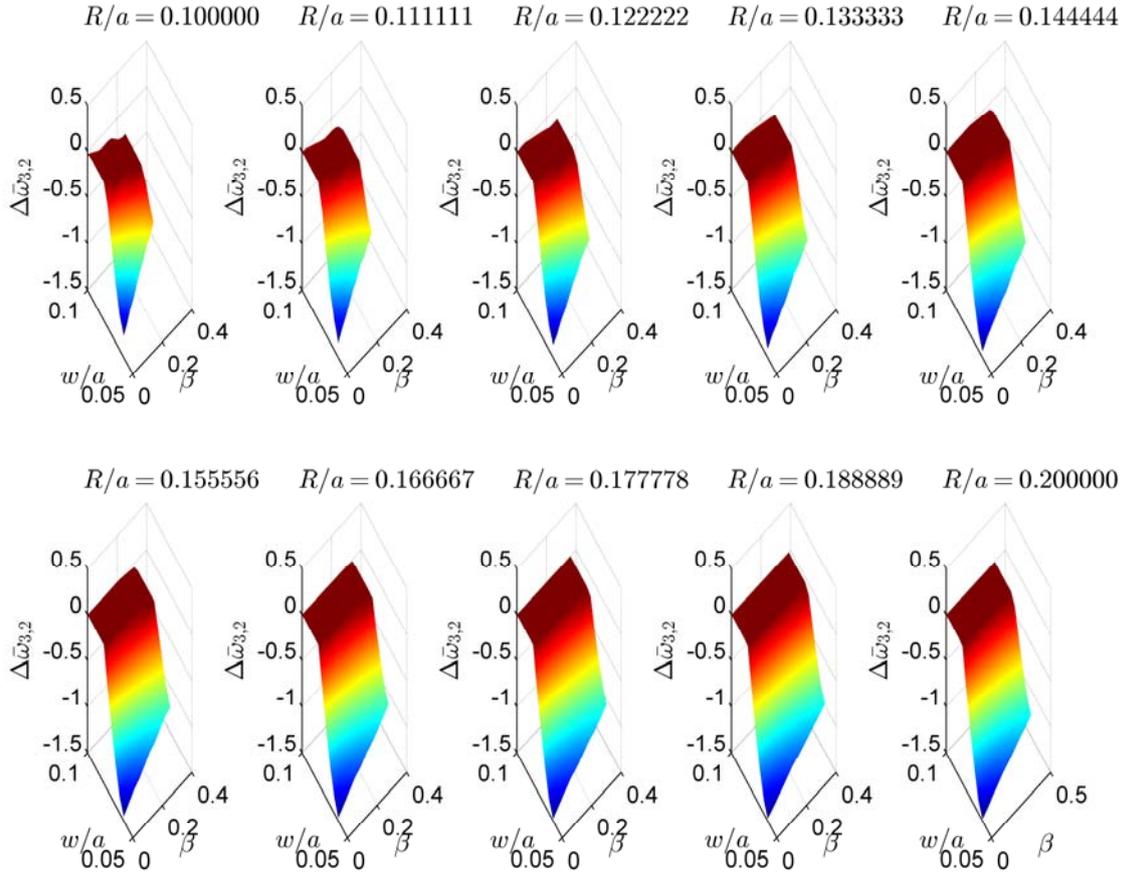

Figure 3: Hexachiral lattice without resonator: band gap $\Delta\bar{\omega}_{3,2}$ as a function of $w/a$ and $\beta$, for several values of $R/a$.

Figure 4 shows the first three dispersion curves at the best solution obtained by the brute-force approach in case there is no resonator, denoted as $BF(3)$, since there are only three parameters to optimize (in the figure, the dimensionless wave number denotes the dimensionless curvilinear coordinate along the boundary of the first irreducible Brillouin zone, see Figure 2 c; moreover, without the resonator, the total number of dispersion curves is exactly 3). Such a solution is $\frac{w}{a} = 0.06$, $\frac{R}{a} = 0.1111$, $\beta = 0.1494$ rad (i.e., $\beta = 8.56°$), and the corresponding band gap is $\Delta\bar{\omega}_{32} = 0$. In this and in the following figures, the functions $\bar{\omega}_h(\bar{\mu},\bar{k})$ and $\bar{\omega}_k(\bar{\mu},\bar{k})$ are plotted when varying $\bar{k}$ on $\partial B_H$ with $H = 300$, i.e., at a larger resolution than in the definition of the objective function (25). The choice of a coarser discretization in the latter is due to the need of limiting the computational burden required to solve the optimization problem (24). In the case considered in Figure 4, one can notice a crossing point between the second and third dispersion curves. Interestingly, the obtained result does not mean that the band gap vanishes for all possible choices of the reduced vector of design variables. Indeed, there exist also other choices for which the band gap is negative. Figure 5 refers to the *worst* such choice, i.e., the one that *minimizes* the band gap (obtained still using the brute-force approach described above, with three optimization variables), which is $\frac{w}{a} = \frac{3}{50}$, $\frac{R}{a} = \frac{1}{5}$, $\beta = 0$ rad (i.e., $\beta = 0°$). The associated band gap is $\Delta\bar{\omega}_{32} = -1.4828$. The figure shows that, in this case, starting from the left, one meets at first a crossing point between the second and third dispersion curves, followed by a veering between the first and second dispersion curves, then a second crossing point between the second and third dispersion curves, a second veering between the first and second dispersion curves, and finally another crossing point between the second and third dispersion curves.

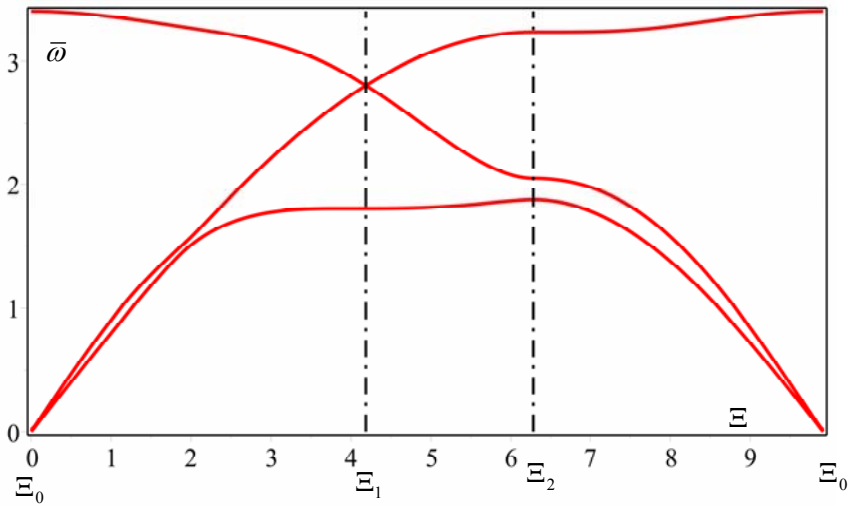

Figure 4: Floquet-Bloch spectrum of the hexachiral lattice without resonator in the boundary of the first irreducible Brillouin zone computed at the best solution found by the brute-force approach.

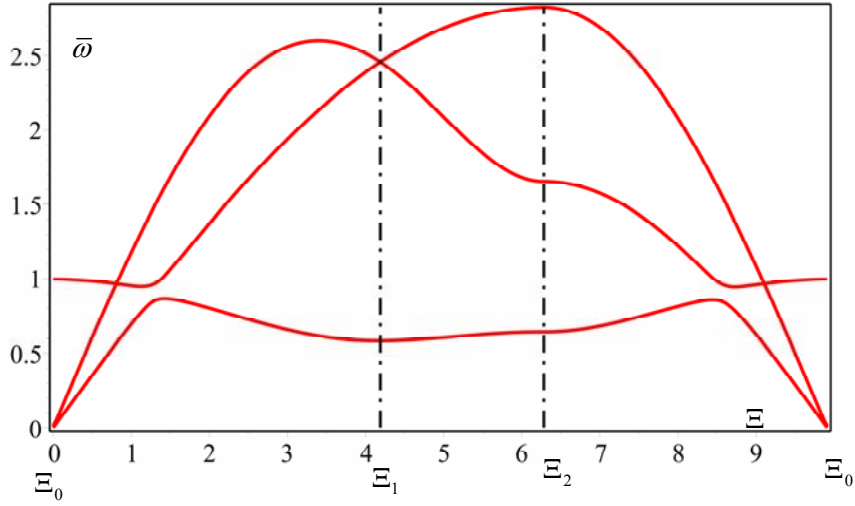

Figure 5: Floquet-Bloch spectrum of the hexachiral lattice without resonator in the boundary of the first irreducible Brillouin zone computed at the worst solution found by the brute-force approach in case there is no resonator.

Next, the case in which all the seven design variables are optimized is considered, corresponding to the situation in which the resonator is introduced in the model. In that case, to apply the brute-force approach, to limit the computational burden, a coarser discretization is needed with respect to the situation above. Figure 6 shows the first three dispersion curves obtained in correspondence of the best choice $\bar{\boldsymbol{\mu}}^{*,BF(7)}$ of the design parameters found by the brute-force approach with seven parameters, using $N_i = 5$ for the design variables $\frac{w}{a}$, $\frac{R}{a}$, $\beta$ (the same parameters optimized in the first numerical experiment above), and $N_i = 3$ for the other four design variables, hence evaluating $5^3 \cdot 2^4 = 2000$ times the objective function $\Delta \bar{\omega}_{32}(\bar{\boldsymbol{\mu}})$. In this case, one obtains

$$\bar{\boldsymbol{\mu}}^{*,BF(7)} = (0.1, 0.1, 0.2014, 0.05, 0.2, 0.1, 10), \tag{29}$$

and the largest value of the objective found by the brute-force approach is

$$\Delta \bar{\omega}_{32}(\bar{\boldsymbol{\mu}}^{*,BF(7)}) = 0.8507, \tag{30}$$

Hence, in this case, there exists a positive band gap. Of course, in such a case, there is no intersection between the second and third dispersion curves.

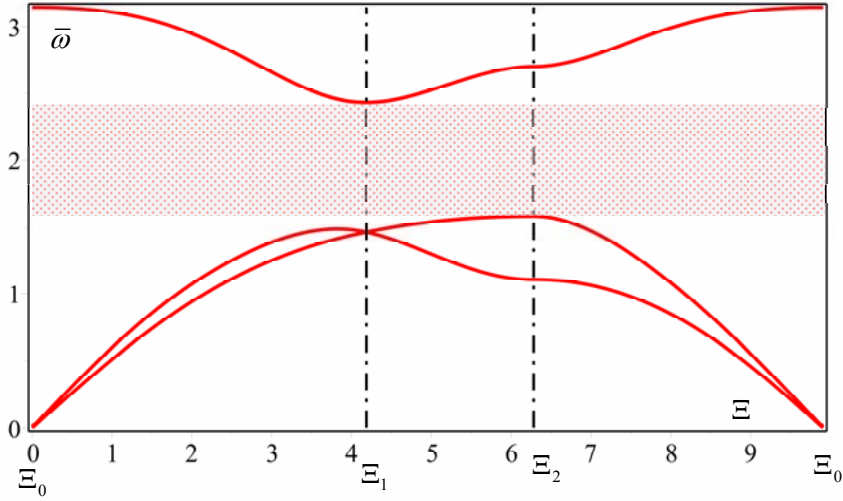

Figure 6: first three dispersion curves of the Floquet-Bloch spectrum of the hexachiral lattice with resonators in the boundary of the first irreducible Brillouin zone computed at the best solution found by the brute-force approach.

Figures 7 and 8 illustrate, instead, for each repetition, the evolution in time of the value of the objective function during the iterations of the adopted optimization method, i.e., the GCMMA algorithm combined with a suitable multi-start technique. In the figures, $m = 0$ refers to the initialization, where a positive value for $m$ refers to the solution obtained in the corresponding iteration of GCMMA. One can also notice that some plots in the two figures are not distinguishable because they almost overlap, as they correspond to a nearly zero-valued objective. The first figure refers to the Monte Carlo initialization, the second one to the quasi-Monte Carlo initialization. One can notice that, during the first iterations, the objective tends sometimes to decrease rather than increase. This is likely due to the fact that, after the initialization of the GCMMA algorithm, the parameters $L_i^{(m)}$, $a_i^{(m)}$, $b_i^{(m)}$, $U_i^{(m)}$ in formula (27) need some iterations before reaching proper values. The solution produced as output by the adopted optimization method (denoted by $\bar{\boldsymbol{\mu}}^{*,GCMMA/MC}$ for the Monte Carlo initialization, and by $\bar{\boldsymbol{\mu}}^{*,GCMMA/QMC}$ for the quasi-Monte Carlo initialization) is the best of all solutions found in this process, i.e., the one associated with the largest value of the objective. In the simulations, the number of iterations in each repetition has been fixed to $M = 24$ (excluding the initialization), and $S = 10$ repetitions have been considered. So, a total of 25 choices for the vector of design variables has been generated in each repetition. The best solution found by the repetitions of the GCMMA algorithm combined with the Monte Carlo initialization is

$$\overline{\boldsymbol{\mu}}^{*,GCMMA/MC} = (0.1, 0.1, 0.1992, 0.052, 0.22, 0.1, 10), \tag{31}$$

and its objective value is

$$\Delta\overline{\omega}_{32}\left(\overline{\boldsymbol{\mu}}^{*,GCMMA/MC}\right) = 0.8596, \tag{32}$$

whereas the best solution found by the GCMMA algorithm combined with the quasi-Monte Carlo initialization is

$$\overline{\boldsymbol{\mu}}^{*,GCMMA/QMC} = (0.1, 0.1, 0.1991, 0.052, 0.22, 0.1, 10), \tag{33}$$

and its objective value is

$$\Delta\overline{\omega}_{32}\left(\overline{\boldsymbol{\mu}}^{*,GCMMA/QMC}\right) = 0.8597, \tag{34}$$

Hence, in this case, $\overline{\boldsymbol{\mu}}^{*} = \overline{\boldsymbol{\mu}}^{*,GCMMA/QMC}$ is the better of the two solutions, and provides a larger band gap than the one obtained by the brute-force approach with seven design parameters.

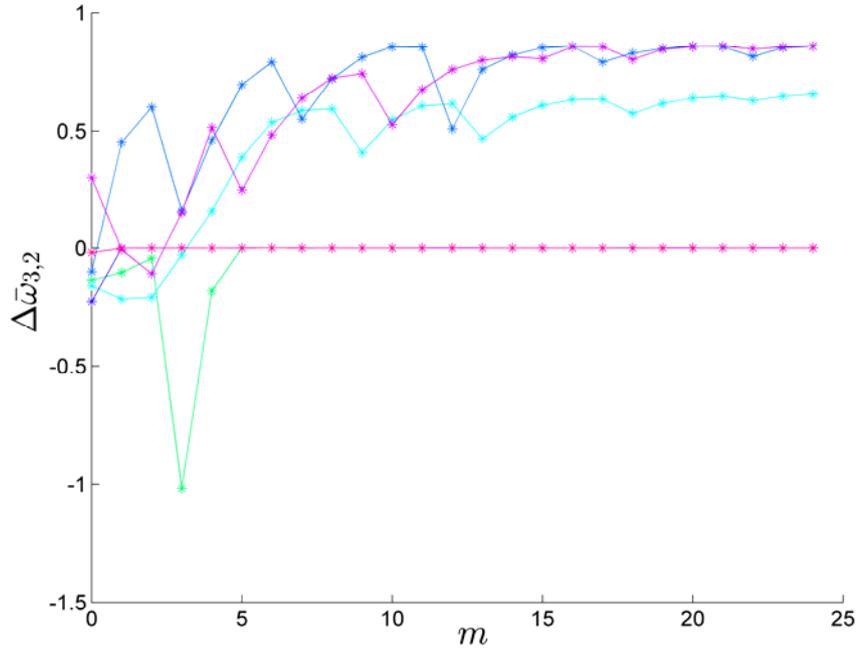

Figure 7: evolution in time of the value of the objective function during the iterations of the GCMMA algorithm combined with the Monte Carlo initialization, for each of the ten repetitions.

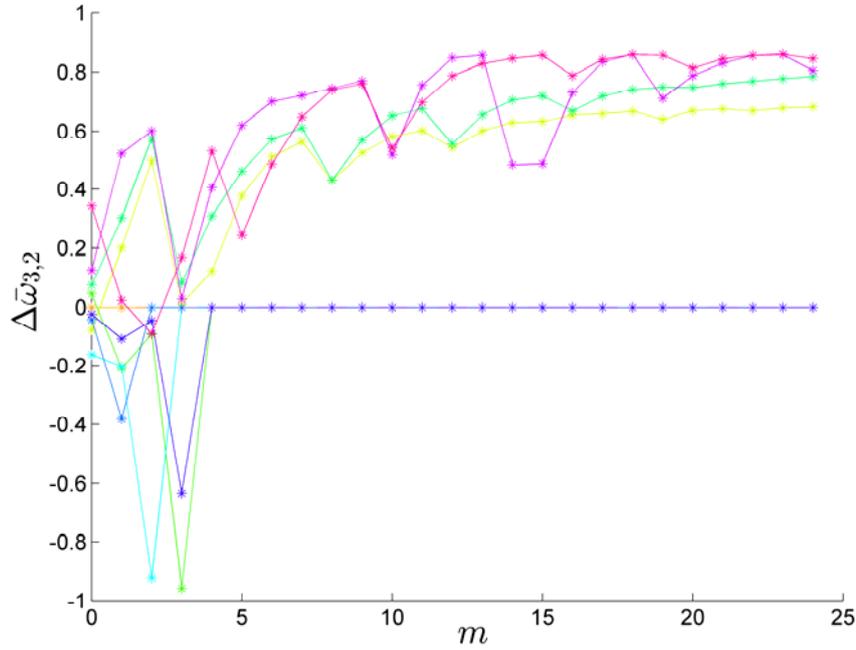

Figure 8: evolution in time of the value of the objective function during the iterations of the GCMMA algorithm combined with the quasi-Monte Carlo initialization, for each of the ten repetitions.

Table 2 reports the best values of the design parameters found by all the optimization methods and considered in the comparison, i.e.: the brute-force approach with three parameters; the brute-force approach with seven parameters; the GCMMA algorithm combined with the Monte Carlo initialization; the GCMMA algorithm combined with the quasi-Monte Carlo initialization.

Table 2: best parameters found by the various optimization methods, and associated objective values

| Method | $w/a$ | $R/a$ | $\beta$ | $r/a$ | $v_a$ | $E_a/E_s$ | $\rho_a/\rho_s$ | $\Delta\bar{\omega}_{32}$ |
|---|---|---|---|---|---|---|---|---|
| BF(3) | 0.06 | 0.1111 | 0.1494 | - | - | - | - | 0 |
| BF(7) | 0.1 | 0.1 | 0.2014 | 0.05 | 0.2 | 0.1 | 10 | 0.8507 |
| GCMMA/MC | 0.1 | 0.1 | 0.1992 | 0.052 | 0.22 | 0.1 | 10 | 0.8596 |
| GCMMA/QMC | 0.1 | 0.1 | 0.1991 | 0.052 | 0.22 | 0.1 | 10 | 0.8597 |

Concluding, both GCMMA/MC and GCMMA/QMC are able to obtain a quite satisfactory approximation of the optimal objective (at least as good as the one obtained by the brute-force approach with seven design parameters) with a much smaller number of objective function evaluations, as compared to the brute-force approach BF(7). Moreover, the best solutions found by all these three methods are similar (see formulas (29), (32), and (33)) but, when compared with the brute-force approach BF(7), the GCMMA algorithm combined with a suitable multi-start technique allows to obtain even larger values of the objective function, since it does not discretize the design variables. Of course, larger values could be also obtained by increasing the number of iterations, still keeping the best solution found in each repetition.

4.2 *Ring-resonator coupling at the best solution found by the adopted optimization method*
In the following, the best solution $\bar{\boldsymbol{\mu}}^* = \bar{\boldsymbol{\mu}}^{*,GCMMA/QMC}$ found by the adopted optimization method is investigated in more detail. Figure 9 shows all its associated 6 dispersion curves. Here, the coupling between the ring and the resonator is evidenced from the fact that there are no horizontal dispersion curves, which would have been obtained in the presence of the resonator only (or of a resonator completely decoupled from the ring). Moreover, this coupling is also evidenced by the presence, in the same unit-norm eigenvectors (using the Euclidean norm on the vector space $\mathbb{C}^6$ to do the normalization, where $\mathbb{C}$ denotes the set of complex numbers), of components related to both the ring and the resonator, in case of dimensionless wave vectors $\bar{\mathbf{k}}$ for which the non-negative dimensionless angular eigenfrequencies are separated, ensuring that each eigenvalue has unit multiplicity. Figure 10 reports all such eigenfrequencies, evaluated on the whole hexagonal Brillouin zone. Finally, Figures 11 and 12 report the components of the unit-norm eigenvectors associated, respectively, with the second and third dispersion surface (such surfaces are obtained by solving the eigenvalue problem (21) on the whole hexagonal Brillouin zone), demonstrating the coupling just discussed, for both the translational and rotational components of the motion of the ring and of the resonator. In particular, the first three components are associated with degree-of-freedom of the ring and the last three are associated with those of the resonator.

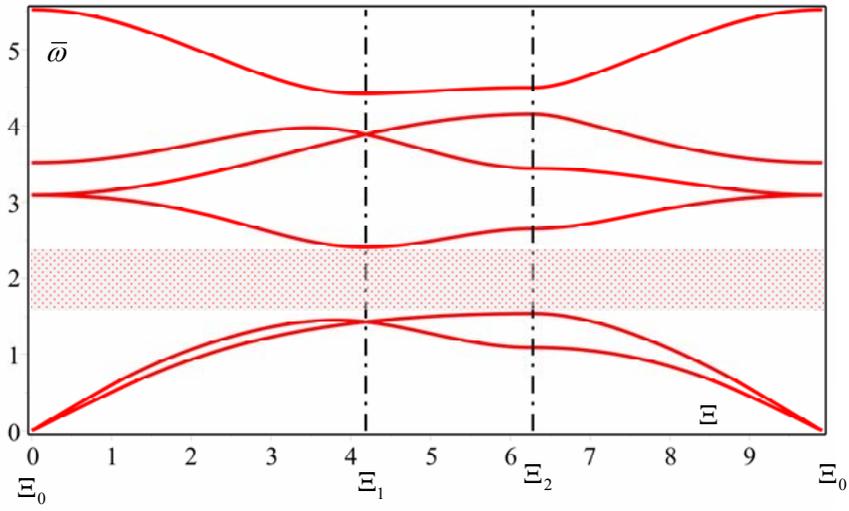

Figure 9: Floquet-Bloch spectrum of the hexachiral lattice with resonators in the boundary of the first irreducible Brillouin zone computed at the best solution found by the GCMMA algorithm combined with the quasi-Monte Carlo initialization.

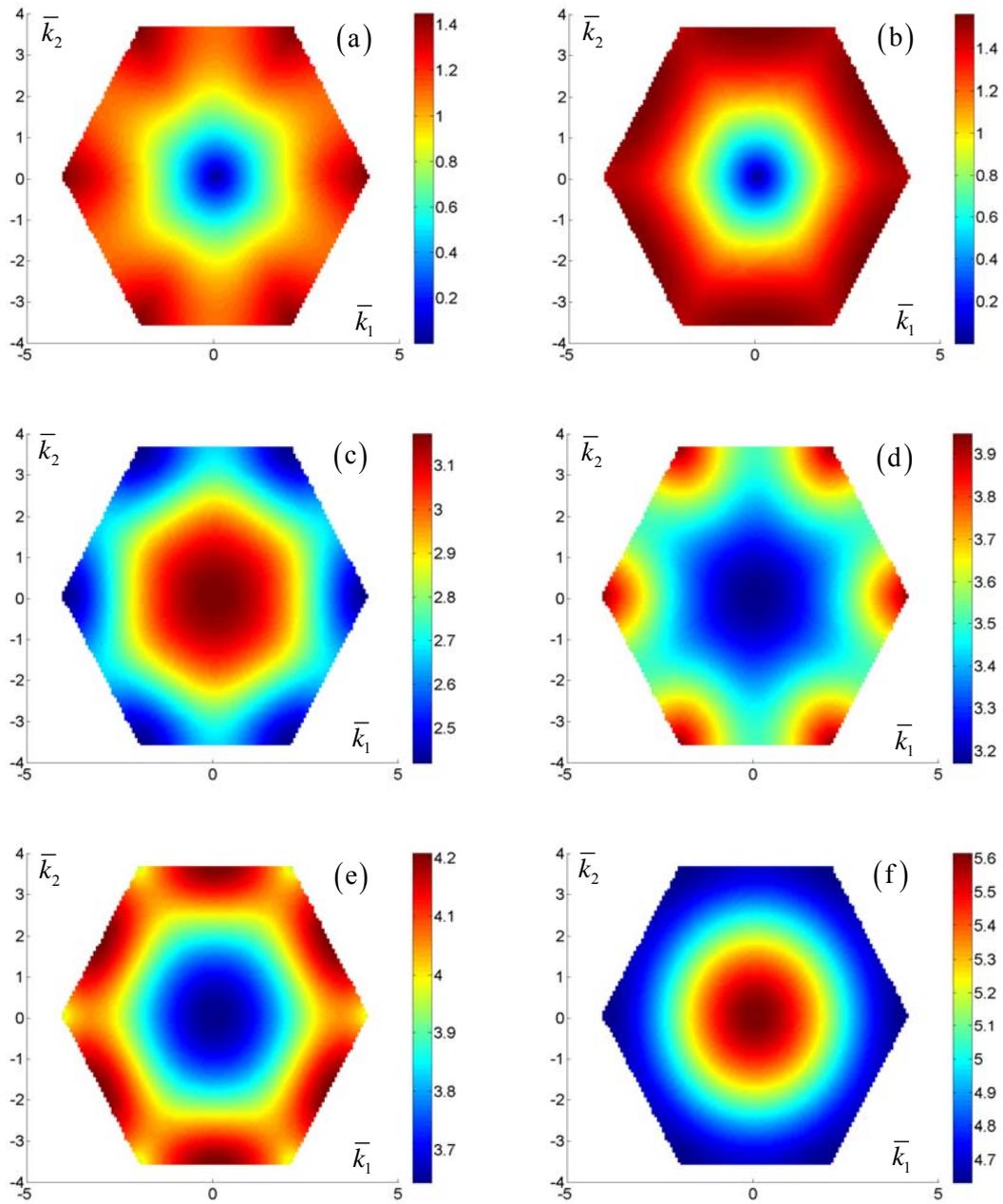

Figure 10: Floquet-Bloch spectrum of the hexachiral lattice with resonators in the hexagonal Brillouin zone computed for the best solution found by the GCMMA algorithm combined with the quasi-Monte Carlo initialization. (a) 1$^{st}$ eigenvalue; (b) 2$^{nd}$ eigenvalue; (c) 3$^{rd}$ eigenvalue; (d) 4$^{th}$ eigenvalue; (e) 5$^{th}$ eigenvalue; (f) 6$^{th}$ eigenvalue.

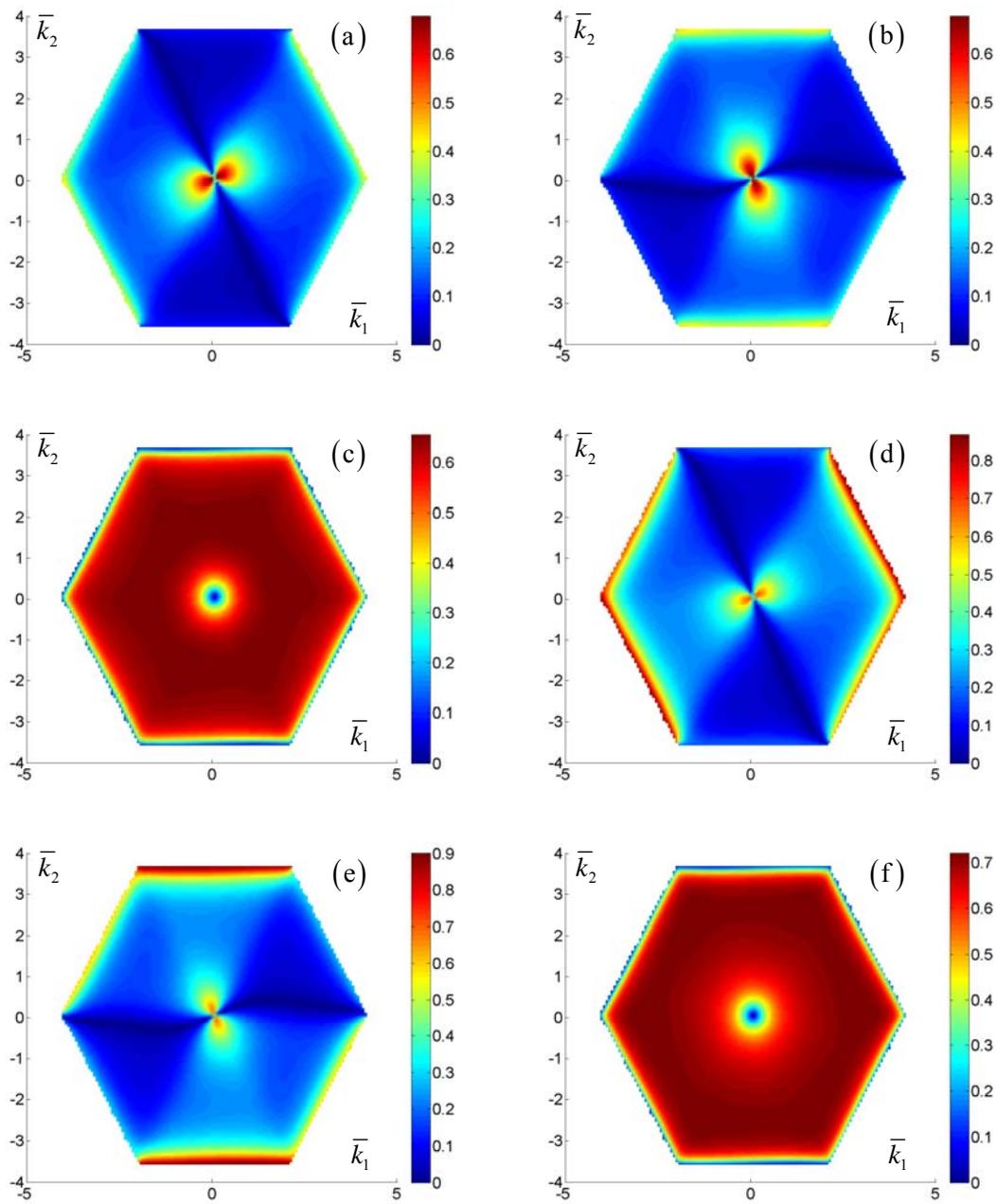

Figure 11: absolute values of the components of the eigenvector associated at the second dispersion surface, computed at the best solution found by the GCMMA algorithm combined with the quasi-Monte Carlo initialization. (a) 1st component; (b) 2nd component; (c) 3rd component; (d) 4th component; (e) 5th component; (f) 6th component.

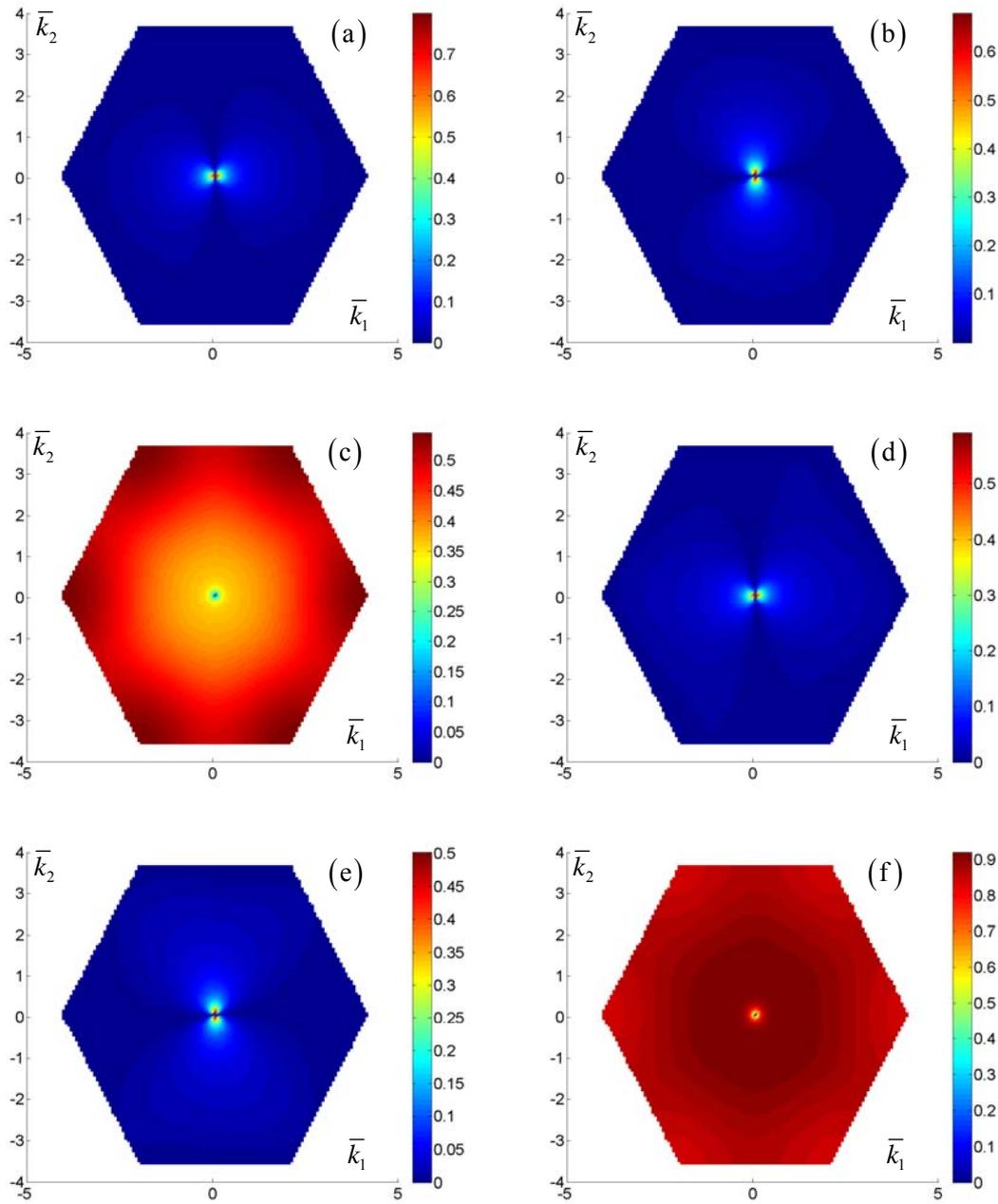

Figure 12: absolute values of the components of the eigenvector associated with the third dispersion surface, computed at the best solution found by the GCMMA algorithm combined with the quasi-Monte Carlo initialization. (a) $1^{st}$ component; (b) $2^{nd}$ component; (c) $3^{rd}$ component; (d) $4^{th}$ component; (e) $5^{th}$ component; (f) $6^{th}$ component.

## 5  Generalizations of the band gap optimization problem

The optimization problem (24) can be extended in various ways. First, instead than maximizing the band gap between two consecutive dispersion curves, one can maximize the weighted sum (with positive weights) of the band gaps associated with several pairs of consecutive dispersion curves, or maximize the minimum of such band gaps. These alternative formulations are suitable for the multiple optimization in the presence of one or more target band gaps. Moreover, in all the formulations of the band gap optimization problem, one can replace each band gap

$$\max_{\bar{\mathbf{k}} \in \partial B_H} \bar{\omega}_h(\bar{\boldsymbol{\mu}}, \bar{\mathbf{k}}) - \min_{\bar{\mathbf{k}} \in \partial B_H} \bar{\omega}_k(\bar{\boldsymbol{\mu}}, \bar{\mathbf{k}}), \tag{35}$$

with the associated *relative* band gap, which is defined as the ratio

$$\frac{\max_{\bar{\mathbf{k}} \in \partial B_H} \bar{\omega}_h(\bar{\boldsymbol{\mu}}, \bar{\mathbf{k}}) - \min_{\bar{\mathbf{k}} \in \partial B_H} \bar{\omega}_k(\bar{\boldsymbol{\mu}}, \bar{\mathbf{k}})}{\frac{1}{2}\left[\min_{\bar{\mathbf{k}} \in \partial B_H} \bar{\omega}_h(\bar{\boldsymbol{\mu}}, \bar{\mathbf{k}}) + \max_{\bar{\mathbf{k}} \in \partial B_H} \bar{\omega}_k(\bar{\boldsymbol{\mu}}, \bar{\mathbf{k}})\right]}, \tag{36}$$

in which, given the same gap amplitude, low-frequency band gaps have higher weight. Such an approach has been considered, e.g., in Men *et al.* 2010 and Men *et al.* 2013.

Another generalization deals with the presence in the objective function of an additional term, related to the *robustness* of the optimal choice of the design variables. Indeed, close to an optimal solution of the (generalized) optimization, the objective function may exhibit high local sensitivity to the design parameter changes. This particular case may occur if the optimal solution lies on the boundary of the parameter domain, as far as the first-order optimality condition cannot be applied in its unconstrained form, so the gradient of the objective function (in case of its local differentiability) is not necessarily a vector of all zeros. Therefore, it may be preferable to look for a solution with a smaller value of the objective, but less sensitive to changes in the design variables. To reach this goal, one can modify the objective function of the problem (1.1), adding a *robustness* term, such as

$$\eta \|\nabla f(\boldsymbol{\mu})\|_2, \tag{37}$$

where $\eta > 0$ is an upper bound on the maximum admissible variation of the Euclidean norm of the vector of design variables with respect to its nominal value. A similar idea was considered in Men *et al.* 2014.

As final remark, the band gap optimization problem and its generalizations has been tackled with different techniques. In some works (for instance, in Huang *et al.* 2014), suitable evolutionary algorithms were used to the purpose. Besides the already mentioned sequential linear programming and CONLIN, one could also make use of the recently proposed method by Men *et al.* 2010 (which replaces the original optimization problem with a sequence of semidefinite programs), and of its extension by Men *et al.* 2013 (which approximates each such semidefinite program with a linear program).

## 6 Conclusions

A parametric beam lattice model has been formulated to analyse the propagation properties of elastic in-plane waves in an auxetic material based on a hexachiral topology of the periodic cell. The material micro-structure is characterized by an ordered assembly of stiff rings connected by flexible ligaments, in the absence of a soft embedding matrix. Inter-ring inclusions are described as linear undamped oscillators and functioning as inertial local resonators, and realize a highly-performant meta-material. A reduced order model in the only dynamically active degrees-of-freedom has been obtained thought the quasi-static condensation of the passive degrees-of-freedom at the periodic cell boundary, where the Floquet-Bloch conditions have been imposed. It has been verified how the introduction of the resonators, if properly tuned, may significantly alter the frequently band structure. From a design perspective, the desirable opening and shifting of band gaps can be obtained in the low-frequency range, paving the way for the realization of passive acoustic filters.

As design testbed, a large space of design geometrical and mechanical parameters has been explored with the objective of opening the largest possible global band gap between the second acoustic branch and the first optical branch in the Floquet-Bloch spectrum. The amplitude maximization has been sought for by the statement of an optimization problem. The solution approach has been based on the Globally Convergent Method of Moving Asymptotes, combined with two suitable multi-start techniques, selected to improve the quality of the obtained locally optimal solution. The method has also been compared with a second, bruce-force approach, showing the larger effectiveness of the former. Finally, the coupling between the ring and the resonator at the best solution found by the adopted optimization method has been investigated. From a qualitative viewpoint, the main results show that, for a periodic cell with fixed characteristic length, the searched high-amplitude band gap can be obtained for small-radius rings and highly-slender, inclined but non-tangent ligaments. Correspondingly, the optimized resonators are found to possess half the radius of the rings and be embedded in a highly-soft matrix.

**Appendix A.1 – Inertial properties and translational and rotational stiffness of the local resonator**

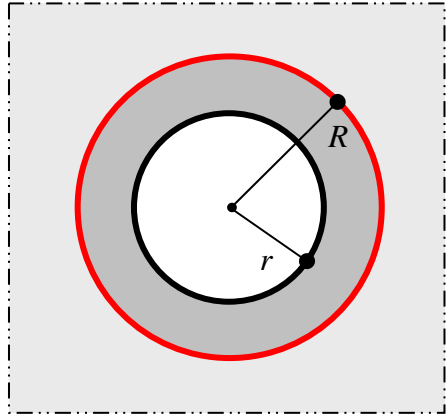

Figure A-1. Rigid disk contained in a soft elastic and isotropic annulus inserted into an external rigid body.

The translational and the rotatory inertia of the rings and the resonator in terms of the geometrical and mechanical properties are $M_s = 2\pi\rho_s R w$, $J_s = M_s R^2$ and $M_r = \pi\rho_r r^2$ and $J_r = \frac{1}{2} M_r r^2$, respectively.

According with Bacigalupo and Gambarotta 2015, the translational and rotational stiffness of the resonator shown in Figure A-1 is here derived. Let us consider first the translation $u$ under plane stress conditions of the rigid disk having radius $r$ surrounded by a homogeneous, elastic, isotropic annulus with Young's modulus $E_r$ and Poisson's ratio $v_r$, having external radius $R$ (Figure A-1). The translational stiffness of the inner disk is evaluated through a FEM analysis by applying a distribution of forces with resultant $F$ to the internal disk. From the displacement $u$, coaxial with $F$, the stiffness $k_d = F/u$ is derived. In Figure A-2a the dimensionless translational stiffness $k_d/E_r$ as a function of the ratio $R/r$ is diagrammatically shown for different values of the Poisson's ratio $v_r$, i.e. $k_d/E_r = f_d(R/r, v_r)$. Therefore, the dimensionless translational stiffness $k_d/E_s$ in terms of ratios $R/r$, $E_r/E_s$, $v_r$ takes the form

$$\frac{k_d}{E_s} = f_d(R/r, v_r) \frac{E_r}{E_s}. \qquad (38)$$

The rotation $\vartheta$ of the rigid inner disk without translation $u$ is analysed by applying to the disk a distribution of forces having resultant torque $M$ and vanishing resultant force $F$. The rotational stiffness is analytically determined in Bacigalupo and Gambarotta 2015 and its dimensionless form in terms of ratios $R/r$, $R/a$, $E_r/E_s$, $\nu_r$ reads

$$\frac{k_\vartheta}{E_s a^2} = \frac{2\pi}{\left((R/r)^2 - 1\right)(1+\nu_r)} \frac{E_r}{E_s} \left(\frac{R}{a}\right)^2 . \qquad (39)$$

In Figure A-2b the dimensionless rotational stiffness $k_\vartheta/E_r r^2$ as a function of the ratio $R/r$ is diagrammatically shown for different values of the Poisson's ratio $\nu_r$.

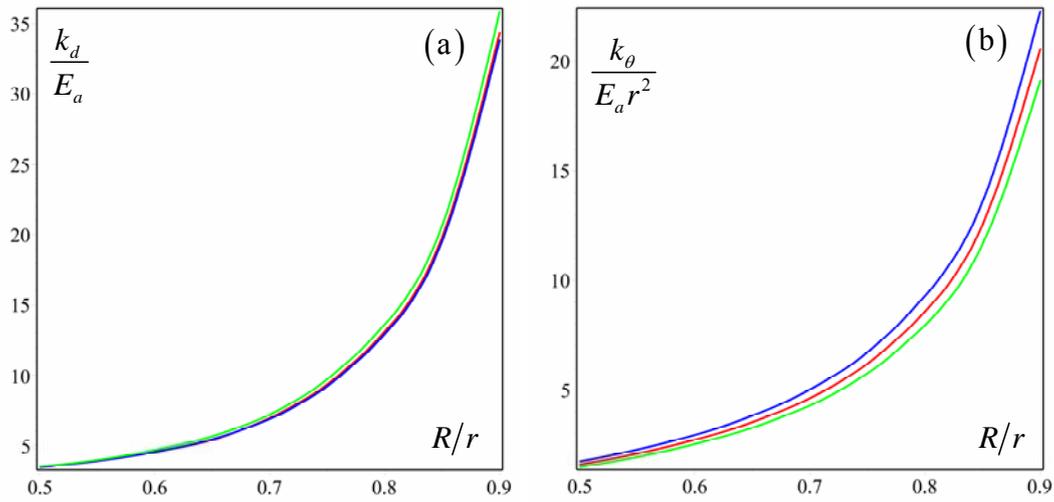

Figure A-2: (a) Dimensionless translational stiffness $k_d/E_r$ in terms of the ratio $R/r$; (b) Dimensionless rotational stiffness $k_\vartheta/E_r R^2$ of the ratio $R/r$. Influence of the Poisson ratio $\nu_r$: blue line $\nu_a = 0.2$; red line $\nu_a = 0.3$; green line $\nu_a = 0.4$.

**Appendix A.2 – Components of the dimensionless matrices $\bar{\mathbf{M}}$ and $\bar{\mathbf{K}}$**

The non vanishing components of the three-by-three positive definite diagonal submatrices $\bar{\mathbf{M}}^s$ and $\bar{\mathbf{M}}^r$, that make up the 6-by-6 dimensionless block diagonal matrix $\bar{\mathbf{M}}(\bar{\boldsymbol{\mu}})$ (see equation (23)), are expressed in terms of the dimensionless parameter vector $\bar{\boldsymbol{\mu}}$ and take the following forms

$$\bar{M}_{11}^s = 2\pi \frac{R}{a}\frac{w}{a}, \quad \bar{M}_{22}^s = 2\pi \frac{R}{a}\frac{w}{a}, \quad \bar{M}_{33}^s = \frac{1}{2}\left(1-\cos^2(\beta)+\Psi^2\right)\pi\frac{R}{a}\frac{w}{a},$$
$$\bar{M}_{11}^r = \pi\left(\frac{r}{a}\right)^2 \frac{\rho_r}{\rho_s}, \quad \bar{M}_{22}^r = \pi\left(\frac{r}{a}\right)^2 \frac{\rho_r}{\rho_s}, \quad \bar{M}_{33}^r = \frac{1}{2}\pi\left(\frac{r}{a}\right)^4 \frac{\rho_r}{\rho_s}, \quad (40)$$

where $\Psi = \sqrt{\cos^2(\beta)+4\left(\frac{R}{a}\right)^2 - 1}$.

The non vanishing components of the three-by-three Hermitian submatrix $\hat{\bar{\mathbf{K}}}_{aa}^s$ and the components of the diagonal submatrices $\bar{\mathbf{K}}_{aa}^r$, $\bar{\mathbf{K}}_{aa}^{sr}$, $\bar{\mathbf{K}}_{aa}^{rs}$, that make up the 6-by-6 dimensionless Hermitian matrix $\bar{\mathbf{K}}(\bar{\boldsymbol{\mu}},\bar{\mathbf{k}})$ (see equation (23)), are expressed in terms of the dimensionless wave vector $\bar{\mathbf{k}}$ and the dimensionless parameter vector $\bar{\boldsymbol{\mu}}$. The components of submatrix $\hat{\bar{\mathbf{K}}}_{aa}^s$ take the following forms

$$\hat{\bar{K}}_{aa\_ij}^s = \frac{1}{\Lambda}\left(\hat{\bar{K}}_{aa\_ij}^{s\_3}\left(\frac{w}{a}\right)^3 + \hat{\bar{K}}_{aa\_ij}^{s\_1}\frac{w}{a} + \hat{\bar{K}}_{aa\_ij}^{s\_0}\right), \quad (41)$$

with $i, j = 1, 2, 3$ being

$$\Lambda = \left(\cos(\beta)-\Psi\right)^3, \quad (42)$$

and the components $\hat{\bar{K}}_{aa\_ij}^{s\_3}$, $\hat{\bar{K}}_{aa\_ij}^{s\_1}$, $\hat{\bar{K}}_{aa\_ij}^{s\_0}$ are defined as follows

$$\hat{\bar{K}}^{s\_3}_{aa\_11} = \left(-2\cos\left(\frac{\sqrt{3}}{2}\bar{k}_2\right)\cos\left(\frac{\bar{k}_1}{2}\right) + 2\cos(\bar{k}_1)\right)\cos^2(\beta) - 2\cos(\bar{k}_1) + 3$$
$$- 2\sqrt{3}\sin\left(\frac{\sqrt{3}}{2}\bar{k}_2\right)\sin\left(\frac{\bar{k}_1}{2}\right)\cos(\beta)\sin(\beta) - \cos\left(\frac{\sqrt{3}}{2}\bar{k}_2\right)\cos\left(\frac{\bar{k}_1}{2}\right),$$

$$\hat{\bar{K}}^{s\_1}_{aa\_11} = \left(2\cos\left(\frac{\sqrt{3}}{2}\bar{k}_2\right)\cos\left(\frac{\bar{k}_1}{2}\right) - 2\cos(\bar{k}_1)\right)\cos^4(\beta)$$
$$+ \left(2\sqrt{3}\sin\left(\frac{\sqrt{3}}{2}\bar{k}_2\right)\sin\left(\frac{\bar{k}_1}{2}\right)\sin(\beta) - 4\Psi\left(\cos\left(\frac{\sqrt{3}}{2}\bar{k}_2\right)\cos\left(\frac{\bar{k}_1}{2}\right) - \cos(\bar{k}_1)\right)\right)\cos^3(\beta)$$
$$+ \left(-4\sqrt{3}\Psi\sin\left(\frac{\sqrt{3}}{2}\bar{k}_2\right)\sin\left(\frac{\bar{k}_1}{2}\right)\sin(\beta) + (2\Psi^2 - 3)\cos\left(\frac{\sqrt{3}}{2}\bar{k}_2\right)\cos\left(\frac{\bar{k}_1}{2}\right) - 2\Psi^2\cos(\bar{k}_1) + 3\right)\cos^2(\beta)$$
$$+ \Psi\left(2\sqrt{3}\Psi\sin\left(\frac{\sqrt{3}}{2}\bar{k}_2\right)\sin\left(\frac{\bar{k}_1}{2}\right)\sin(\beta) + 6\cos\left(\frac{\sqrt{3}}{2}\bar{k}_2\right)\cos\left(\frac{\bar{k}_1}{2}\right) - 6\right)\cos(\beta)$$
$$- 3\Psi^2\left(\cos\left(\frac{\sqrt{3}}{2}\bar{k}_2\right)\cos\left(\frac{\bar{k}_1}{2}\right) - 1\right),$$

$$\hat{\bar{K}}^{s\_0}_{aa\_11} = \frac{k_d}{E_s}\cos^3(\beta) - 3\Psi\frac{k_d}{E_s}\cos^2(\beta) + 3\Psi^2\frac{k_d}{E_s}\cos(\beta) - \Psi^3\frac{k_d}{E_s}, \tag{43}$$

$$\hat{\bar{K}}^{s\_3}_{aa\_22} = \left(2\cos\left(\frac{\sqrt{3}}{2}\bar{k}_2\right)\cos\left(\frac{\bar{k}_1}{2}\right) - 2\cos(\bar{k}_1)\right)\cos^2(\beta) + 3$$
$$+ 2\sqrt{3}\sin\left(\frac{\sqrt{3}}{2}\bar{k}_2\right)\sin\left(\frac{\bar{k}_1}{2}\right)\cos(\beta)\sin(\beta) - 3\cos\left(\frac{\sqrt{3}}{2}\bar{k}_2\right)\cos\left(\frac{\bar{k}_1}{2}\right),$$

$$\hat{\bar{K}}^{s\_1}_{aa\_22} = \left(-2\cos\left(\frac{\sqrt{3}}{2}\bar{k}_2\right)\cos\left(\frac{\bar{k}_1}{2}\right) + 2\cos(\bar{k}_1)\right)\cos^4(\beta)$$
$$+ \left(-2\sqrt{3}\sin\left(\frac{\sqrt{3}}{2}\bar{k}_2\right)\sin\left(\frac{\bar{k}_1}{2}\right)\sin(\beta) + 4\Psi\left(\cos\left(\frac{\sqrt{3}}{2}\bar{k}_2\right)\cos\left(\frac{\bar{k}_1}{2}\right) - \cos(\bar{k}_1)\right)\right)\cos^3(\beta)$$
$$+ \left(4\sqrt{3}\Psi\sin\left(\frac{\sqrt{3}}{2}\bar{k}_2\right)\sin\left(\frac{\bar{k}_1}{2}\right)\sin(\beta) - (2\Psi^2 + 1)\cos\left(\frac{\sqrt{3}}{2}\bar{k}_2\right)\cos\left(\frac{\bar{k}_1}{2}\right) + 2(\Psi^2 - 1)\cos(\bar{k}_1) + 3\right)\cos^2(\beta)$$
$$+ \Psi\left(-2\sqrt{3}\Psi\sin\left(\frac{\sqrt{3}}{2}\bar{k}_2\right)\sin\left(\frac{\bar{k}_1}{2}\right)\sin(\beta) + 2\cos\left(\frac{\sqrt{3}}{2}\bar{k}_2\right)\cos\left(\frac{\bar{k}_1}{2}\right) + 4\cos(\bar{k}_1) - 6\right)\cos(\beta)$$
$$+ \Psi^2\left(3 - \cos\left(\frac{\sqrt{3}}{2}\bar{k}_2\right)\cos\left(\frac{\bar{k}_1}{2}\right) - 2\cos(\bar{k}_1)\right),$$

$$\hat{\bar{K}}^{s\_0}_{aa\_22} = \frac{k_d}{E_s}\cos^3(\beta) - 3\Psi\frac{k_d}{E_s}\cos^2(\beta) + 3\Psi^2\frac{k_d}{E_s}\cos(\beta) - \Psi^3\frac{k_d}{E_s}, \tag{44}$$

$$\hat{\bar{K}}^{s\_3}_{aa\_33} = \left(\frac{2}{3}\cos\left(\frac{\sqrt{3}}{2}\bar{k}_2\right)\cos\left(\frac{\bar{k}_1}{2}\right) + \frac{2}{3}\cos(\bar{k}_1) + 2\right)\cos^2(\beta) + \frac{1}{6}\Psi^2\cos(\bar{k}_1)$$

$$+ \left(\frac{2}{3}\cos\left(\frac{\sqrt{3}}{2}\bar{k}_2\right)\cos\left(\frac{\bar{k}_1}{2}\right) + \frac{1}{3}\Psi\cos(\bar{k}_1) - \Psi\right)\cos(\beta)$$

$$+ \Psi^2\left(\frac{1}{2} - \frac{1}{3}\cos\left(\frac{\sqrt{3}}{2}\bar{k}_2\right)\cos\left(\frac{\bar{k}_1}{2}\right)\right),$$

$$\hat{\bar{K}}^{s\_1}_{aa\_33} = \left(-\cos\left(\frac{\sqrt{3}}{2}\bar{k}_2\right)\cos\left(\frac{\bar{k}_1}{2}\right) - \frac{1}{2}\cos(\bar{k}_1) - \frac{3}{2}\right)\cos^4(\beta) + \Psi\left(2\cos\left(\frac{\sqrt{3}}{2}\bar{k}_2\right)\cos\left(\frac{\bar{k}_1}{2}\right) + \cos(\bar{k}_1) + 3\right)\cos^3(\beta)$$

$$+ \left(1 - \Psi^2\right)\left(\cos\left(\frac{\sqrt{3}}{2}\bar{k}_2\right)\cos\left(\frac{\bar{k}_1}{2}\right) + \frac{3}{2} + \frac{1}{2}\cos(\bar{k}_1)\right)\cos^2(\beta)$$

$$+ \Psi\left(-2\cos\left(\frac{\sqrt{3}}{2}\bar{k}_2\right)\cos\left(\frac{\bar{k}_1}{2}\right) - \cos(\bar{k}_1) - 3\right)\cos(\beta) + \Psi^2\left(\frac{1}{2}\cos(\bar{k}_1) + \frac{3}{2} + \cos\left(\frac{\sqrt{3}}{2}\bar{k}_2\right)\cos\left(\frac{\bar{k}_1}{2}\right)\right),$$

$$\hat{\bar{K}}^{s\_0}_{aa\_33} = \frac{k_g}{a^2 E_s}\left(3\Psi^2\cos(\beta) - 3\Psi\cos^2(\beta) + \cos^3(\beta) - \Psi^3\right),$$

(45)

$$\hat{\bar{K}}^{s\_3}_{aa\_12} = -2\sqrt{3}\sin\left(\frac{\sqrt{3}}{2}\bar{k}_2\right)\sin\left(\frac{\bar{k}_1}{2}\right)\cos^2(\beta) + \sqrt{3}\sin\left(\frac{\sqrt{3}}{2}\bar{k}_2\right)\sin\left(\frac{\bar{k}_1}{2}\right)$$

$$+ \left(2\cos\left(\frac{\sqrt{3}}{2}\bar{k}_2\right)\cos\left(\frac{\bar{k}_1}{2}\right) - 2\cos(\bar{k}_1)\right)\cos(\beta)\sin(\beta),$$

$$\hat{\bar{K}}^{s\_1}_{aa\_12} = 2\sqrt{3}\sin\left(\frac{\sqrt{3}}{2}\bar{k}_2\right)\sin\left(\frac{\bar{k}_1}{2}\right)\cos^4(\beta) - \sqrt{3}\Psi^2\sin\left(\frac{\sqrt{3}}{2}\bar{k}_2\right)\sin\left(\frac{\bar{k}_1}{2}\right)$$

$$+ \left(-4\sqrt{3}\Psi\sin\left(\frac{\sqrt{3}}{2}\bar{k}_2\right)\sin\left(\frac{\bar{k}_1}{2}\right) + \left(2\cos(\bar{k}_1) - 2\cos\left(\frac{\sqrt{3}}{2}\bar{k}_2\right)\cos\left(\frac{\bar{k}_1}{2}\right)\right)\sin(\beta)\right)\cos^3(\beta) \quad (46)$$

$$+ \left(\sqrt{3}\left(2\Psi^2 - 1\right)\sin\left(\frac{\sqrt{3}}{2}\bar{k}_2\right)\sin\left(\frac{\bar{k}_1}{2}\right) + 4\Psi\left(\cos\left(\frac{\sqrt{3}}{2}\bar{k}_2\right)\cos\left(\frac{\bar{k}_1}{2}\right) - \cos(\bar{k}_1)\right)\sin(\beta)\right)\cos^2(\beta)$$

$$+ \Psi\left(2\Psi\left(\cos(\bar{k}_1) - \cos\left(\frac{\sqrt{3}}{2}\bar{k}_2\right)\cos\left(\frac{\bar{k}_1}{2}\right)\right)\sin(\beta) + 2\sqrt{3}\sin\left(\frac{\sqrt{3}}{2}\bar{k}_2\right)\sin\left(\frac{\bar{k}_1}{2}\right)\right)\cos(\beta),$$

$$\hat{\bar{K}}^{s\_0}_{aa\_12} = 0,$$

$$\hat{\bar{K}}^{s\_3}_{aa\_13} = -i\sqrt{3}\sin\left(\frac{\sqrt{3}}{2}\bar{k}_2\right)\cos\left(\frac{\bar{k}_1}{2}\right)\cos^2(\beta)$$

$$+ i\left(\cos\left(\frac{\sqrt{3}}{2}\bar{k}_2\right)\sin\left(\frac{\bar{k}_1}{2}\right) + \sin(\bar{k}_1)\right)\cos(\beta)\sin(\beta),$$

$$\hat{\bar{K}}^{s\_1}_{aa\_13} = -i\sqrt{3}\sin\left(\frac{\sqrt{3}}{2}\bar{k}_2\right)\cos\left(\frac{\bar{k}_1}{2}\right)\cos^4(\beta) + i\sqrt{3}\Psi^2\sin\left(\frac{\sqrt{3}}{2}\bar{k}_2\right)\cos\left(\frac{\bar{k}_1}{2}\right)$$

$$+ i\left(2\sqrt{3}\Psi\sin\left(\frac{\sqrt{3}}{2}\bar{k}_2\right)\cos\left(\frac{\bar{k}_1}{2}\right) + \left(\sin(\bar{k}_1) + \cos\left(\frac{\sqrt{3}}{2}\bar{k}_2\right)\sin\left(\frac{\bar{k}_1}{2}\right)\right)\sin(\beta)\right)\cos^3(\beta) \quad (47)$$

$$+ i\left(\sqrt{3}(1-\Psi^2)\sin\left(\frac{\sqrt{3}}{2}\bar{k}_2\right)\cos\left(\frac{\bar{k}_1}{2}\right) - 2\Psi\left(\cos\left(\frac{\sqrt{3}}{2}\bar{k}_2\right)\sin\left(\frac{\bar{k}_1}{2}\right) + \sin(\bar{k}_1)\right)\sin(\beta)\right)\cos^2(\beta)$$

$$+ i\Psi\left(\Psi\left(\sin(\bar{k}_1) + \cos\left(\frac{\sqrt{3}}{2}\bar{k}_2\right)\sin\left(\frac{\bar{k}_1}{2}\right)\right)\sin(\beta) - 2\sqrt{3}\sin\left(\frac{\sqrt{3}}{2}\bar{k}_2\right)\cos\left(\frac{\bar{k}_1}{2}\right)\right)\cos(\beta),$$

$$\hat{\bar{K}}^{s\_0}_{aa\_13} = 0,$$

$$\hat{\bar{K}}^{s\_3}_{aa\_23} = i\left(\cos\left(\frac{\sqrt{3}}{2}\bar{k}_2\right)\sin\left(\frac{\bar{k}_1}{2}\right) + \sin(\bar{k}_1)\right)\cos^2(\beta)$$

$$+ i\sqrt{3}\sin\left(\frac{\sqrt{3}}{2}\bar{k}_2\right)\cos\left(\frac{\bar{k}_1}{2}\right)\sin(\beta)\cos(\beta),$$

$$\hat{\bar{K}}^{s\_1}_{aa\_23} = i\left(\cos\left(\frac{\sqrt{3}}{2}\bar{k}_2\right)\sin\left(\frac{\bar{k}_1}{2}\right) + \sin(\bar{k}_1)\right)\cos^4(\beta) - \Psi^2\left(\cos\left(\frac{\sqrt{3}}{2}\bar{k}_2\right)\sin\left(\frac{\bar{k}_1}{2}\right) + \sin(\bar{k}_1)\right)$$

$$+ i\left(\sqrt{3}\sin\left(\frac{\sqrt{3}}{2}\bar{k}_2\right)\cos\left(\frac{\bar{k}_1}{2}\right)\sin(\beta) - 2\Psi\left(\cos\left(\frac{\sqrt{3}}{2}\bar{k}_2\right)\sin\left(\frac{\bar{k}_1}{2}\right) + \sin(\bar{k}_1)\right)\right)\cos^3(\beta)$$

$$+ i\left((\Psi^2-1)\left(\cos\left(\frac{\sqrt{3}}{2}\bar{k}_2\right)\sin\left(\frac{\bar{k}_1}{2}\right) + \sin(\bar{k}_1)\right) - 2\Psi\sqrt{3}\sin\left(\frac{\sqrt{3}}{2}\bar{k}_2\right)\cos\left(\frac{\bar{k}_1}{2}\right)\sin(\beta)\right)\cos^2(\beta)$$

$$+ i\Psi\left(\sqrt{3}\Psi\sin\left(\frac{\sqrt{3}}{2}\bar{k}_2\right)\cos\left(\frac{\bar{k}_1}{2}\right)\sin(\beta) + 2\cos\left(\frac{\sqrt{3}}{2}\bar{k}_2\right)\sin\left(\frac{\bar{k}_1}{2}\right) + 2\sin(\bar{k}_1)\right)\cos(\beta),$$

$$\hat{\bar{K}}^{s\_0}_{aa\_23} = 0, \quad (48)$$

$$\hat{\bar{K}}^{s\_n}_{aa\_21} = \hat{\bar{K}}^{s\_n}_{aa\_12}, \qquad \hat{\bar{K}}^{s\_n}_{aa\_31} = -i\,\text{Im}\left(\hat{\bar{K}}^{s\_n}_{aa\_13}\right), \qquad \hat{\bar{K}}^{s\_n}_{aa\_32} = -i\,\text{Im}\left(\hat{\bar{K}}^{s\_n}_{aa\_23}\right), \quad (49)$$

where $\Psi = \sqrt{\cos^2(\beta) + 4\left(\dfrac{R}{a}\right)^2 - 1}$, $n = 0,1,3$, $i^2 = -1$ and $\text{Im}(z)$ denotes the imaginary part of the complex numbers $z$.

Finally, the non vanishing components of the diagonal submatrix $\bar{\mathbf{K}}^r_{aa}$ are

$$\bar{K}^r_{aa\_11} = \frac{k_d}{E_s}, \qquad \bar{K}^r_{aa\_22} = \frac{k_d}{E_s}, \qquad \bar{K}^r_{aa\_33} = \frac{k_\vartheta}{a^2 E_s}, \qquad (50)$$

whereas the diagonal submatrices $\bar{\mathbf{K}}^{sr}_{aa}$, $\bar{\mathbf{K}}^{rs}_{aa}$ satisfy the constraint $\bar{\mathbf{K}}^{sr}_{aa} = (\bar{\mathbf{K}}^{rs}_{aa})^{\mathrm{T}} = -\bar{\mathbf{K}}^r_{aa}$.